\documentstyle[prd,aps,preprint,tighten,epsfig]{revtex}

\newcommand{\beq}{\begin{equation}}
\newcommand{\eeq}{\end{equation}}
\newcommand{\bea}{\begin{eqnarray}}
\newcommand{\eea}{\end{eqnarray}}

\draft

\begin{document}

\title{On the Fukugita-Tanimoto-Yanagida Ansatz with Partially
Non-degenerate Right-handed Majorana Neutrinos}

\author{{\bf Midori Obara} \thanks{E-mail: midori@mail.ihep.ac.cn}
~ and ~ {\bf Zhi-zhong Xing} \thanks{E-mail:
xingzz@mail.ihep.ac.cn}}

\address{
%CCAST (World Laboratory), P.O. Box 8730, Beijing 100080,
%China \\
%and
Institute of High Energy Physics, Chinese Academy of
Sciences, \\
P.O. Box 918, Beijing 100049, China}

\maketitle

\begin{abstract}
Taking three right-handed Majorana neutrino masses $M^{}_i$ to be
partially non-degenerate, we make a new analysis of the
Fukugita-Tanimoto-Yanagida ansatz and confront it with current
neutrino oscillation data. We determine the parameter space for
cases (A) $M^{}_3 = M^{}_2 \neq M^{}_1$, (B) $M^{}_2 = M^{}_1 \neq
M^{}_3$ and (C) $M^{}_1 = M^{}_3 \neq M^{}_2$, and examine their
respective deviations from the original $M^{}_1 = M^{}_2 = M^{}_3$
case. The numerical constraints on three light neutrino masses,
three neutrino mixing angles and three CP-violating phases are
also obtained, together with the predictions for the Jarlskog
invariant of CP violation and the effective masses of the tritium
beta decay and the neutrinoless double-beta decay.
\end{abstract}


\section{Introduction}

Very robust evidence for the existence of neutrino oscillations
has recently been achieved from solar \cite{SNO}, atmospheric
\cite{SK}, reactor \cite{KM} and accelerator \cite{K2K} neutrino
experiments. Thanks to this exciting progress in neutrino physics,
we are now convinced that neutrinos are massive and lepton flavors
are mixed. The phenomenon of lepton flavor mixing can be described
by a $3\times 3$ unitary matrix $V$, the Maki-Nakagawa-Sakata
(MNS) matrix \cite{MNS}, which contains three mixing angles
($\theta^{}_{12}$, $\theta^{}_{23}$, $\theta^{}_{13}$) and three
CP-violating phases ($\delta$, $\rho$, $\sigma$). Four of these
six parameters (i.e., $\theta^{}_{12}$, $\theta^{}_{23}$,
$\theta^{}_{13}$ and $\delta$), together with two neutrino
mass-squared differences ($\Delta m_{21}^2 \equiv m_2^2 - m_1^2$
and $\Delta m_{32}^2 \equiv m_3^2 - m_2^2$), can be extracted from
the measurements of neutrino oscillations. A global analysis of
current experimental data yields \cite{vissani}
\begin{eqnarray}
0.25 < & \sin^2 \theta_{12} ~ & < 0.38 \; , \nonumber \\
0.35 < & \sin^2 \theta_{23} & < 0.65 \; , \nonumber \\
& \sin^2 \theta_{13} & < 0.03 \; ;
%       (1)
\label{theta12}
\end{eqnarray}
and
\begin{eqnarray}
7.2 \times 10^{-5} ~ {\rm eV}^2 \leq & \Delta m_{21}^2 & \leq 8.9
\times
10^{-5} ~ {\rm eV}^2 \; , \nonumber \\
2.1 \times 10^{-3} ~ {\rm eV}^2 \leq & |\Delta m_{32}^2| ~ & \leq
3.1 \times 10^{-5} ~ {\rm eV}^2 \; ,
%       (2)
\label{m31}
\end{eqnarray}
%%%---------
at the $99 \%$ confidence level, but the Dirac CP-violating phase
$\delta$ is entirely unrestricted at present. More accurate
neutrino oscillation experiments are going to determine the size
of $\theta^{}_{13}$, the sign of $\Delta m^2_{32}$ and the
magnitude of $\delta$. The proposed precision experiments for the
tritium beta decay \cite{KATRIN} and the neutrinoless double-beta
decay \cite{0nubeta} will help to probe the absolute mass scale of
three light neutrinos and to constrain the Majorana CP-violating
phases $\rho$ and $\sigma$.

Towards much better understanding of the neutrino mass spectrum
and the neutrino mixing pattern indicated by Eqs. (1) and (2),
many phenomenological ans$\rm\ddot{a}$tze of lepton mass matrices
have recently been considered and discussed \cite{Review}. Among
them, a particularly simple example \cite{Xing02} is to assume
that both the charged-lepton mass matrix $M^{}_l$ and the neutrino
mass matrix $M^{}_\nu$ are of the well-known Fritzsch texture
\cite{F78}
%%%%%%%%%%%%%%%%%%%%%%%%%%%%%%
\footnote{See Ref. \cite{Ago} for some earlier applications of the
Fritzsch ansatz to the lepton sector. In these works, however,
only the small-mixing-angle MSW solution to the solar neutrino
problem was taken into account. This solution is now out of
date.}.
%%%%%%%%%%%%%%%%%%%%%%%%%%%%%%
Incorporating the Fritzsch texture in the canonical seesaw
mechanism \cite{SS} with three degenerate right-handed Majorana
neutrinos, Fukugita, Tanimoto and Yanagida (FTY) \cite{FTY} have
proposed an interesting ansatz to account for current neutrino
oscillation data. To be explicit, the FTY ansatz includes two
assumptions: (1) both the charged-lepton mass matrix $M^{}_l$ and
the Dirac neutrino mass matrix $M^{}_{\rm D}$ take the Fritzsch
texture; and (2) the right-handed Majorana neutrino mass matrix
$M^{}_{\rm R}$ takes the form $M^{}_{\rm R} = M^{}_0 {\bf 1}$ with
$\bf 1$ being the $3\times 3$ unity matrix (i.e., $M^{}_i =
M^{}_0$ for $i=1$, 2 and 3). Then the effective (left-handed)
neutrino mass matrix $M^{}_\nu$ in the FTY ansatz is given by
\begin{equation}
M^{}_\nu \; =\; M^{}_{\rm D} M^{-1}_{\rm R} M^T_{\rm D} \; =\;
\frac{M^2_{\rm D}}{M^{}_0} \; .
%       (3)
\end{equation}
Unless a special assumption is further made \cite{XZ05}, the
texture of $M^{}_\nu$ is no more of the Fritzsch form. Refs.
\cite{FTY} and \cite{ZX05} have shown that the FTY ansatz is
compatible very well with current experimental data on solar and
atmospheric neutrino oscillations.

The main purpose of this work is to go beyond the FTY ansatz by
relaxing its second assumption; namely, the masses of three
right-handed Majorana neutrinos are allowed to be partially
non-degenerate. We consider three different cases: (A) $M^{}_3 =
M^{}_2 \neq M^{}_1$, (B) $M^{}_2 = M^{}_1 \neq M^{}_3$, and (C)
$M^{}_1 = M^{}_3 \neq M^{}_2$. Indeed, the non-degeneracy or
partial non-degeneracy of $M^{}_i$ (for $i=1, 2, 3$) is a
necessary condition to get the successful thermal leptogenesis
\cite{FY} in order to account for the cosmological
baryon-antibaryon asymmetry. See Ref. \cite{XZ06} for an explicit
example which illustrates the correlation between the mass
splitting of heavy right-handed Majorana neutrinos and the
leptogenesis in the minimal seesaw model. Here we focus on the
low-energy phenomenology of the generalized FTY ansatz --- in
particular, we are going to examine how the neutrino masses,
flavor mixing angles and CP-violating phases are sensitive to the
mass splitting parameters in cases (A), (B) and (C), which deviate
from the original $M^{}_1 = M^{}_2 = M^{}_3$ case in different
ways. We shall discuss the possibilities to further extend the FTY
ansatz to accommodate leptogenesis somewhere else.

The remaining part of our paper is organized as follows. In
section II, we make some reasonable analytical approximations to
calculate the neutrino mass matrix $M^{}_\nu$, in which the mass
splitting of any two heavy Majorana neutrinos has been taken into
account. The neutrino mass spectrum and the MNS matrix can then be
derived for each of the three cases. Section III is devoted to a
detailed numerical analysis of the allowed parameter space in each
case, and to the determination of three neutrino masses, three
mixing angles and three CP-violating phases. We also obtain the
predictions for the Jarlskog invariant of CP violation and the
effective masses of the tritium beta decay and the neutrinoless
double-beta decay. Finally, a brief summary of our main results is
given in section IV.

\section{Analytical calculations}

Let us assume that the charged-lepton mass matrix $M^{}_l$ and the
Dirac neutrino mass matrix $M^{}_{\rm D}$ are both symmetric and
of the Fritzsch texture:
\begin{equation}
M^{}_{l(\rm D)} \; = \; \left ( \matrix{ {\bf 0} & ~~ C^{}_{l(\rm
D)} ~~ & {\bf 0} \cr C^{}_{l(\rm D)} & {\bf 0} & B^{}_{l(\rm D)}
\cr {\bf 0} & B^{}_{l(\rm D)} & A^{}_{l(\rm D)} \cr} \right ) \; ,
%       (4)
\end{equation}
in which only $A^{}_{l(\rm D)}$ is real and positive. $M^{}_{l(\rm
D)}$ can be decomposed as $M^{}_{l(\rm D)} = P^{}_{l(\rm D)}
\overline{M}^{}_{l(\rm D)} P_{l(\rm D)}^T$, where $P^{}_{l(\rm D)}
={\rm Diag} \, \{ e^{i[\varphi^{}_{l(\rm D)} - \phi^{}_{l(\rm
D)}]}, e^{i \phi^{}_{l(\rm D)}}, 1\}$ with $\phi^{}_{l(\rm D)}
\equiv \arg [B^{}_{l(\rm D)}]$ and $\varphi^{}_{l(\rm D)} \equiv
\arg [C^{}_{l(\rm D)}]$, and
\begin{equation}
\overline{M}^{}_{l(\rm D)} = \left ( \matrix{ {\bf 0} &
|C^{}_{l(\rm D)}| & {\bf 0} \cr |C^{}_{l(\rm D)}| & {\bf 0} &
|B^{}_{l(\rm D)}| \cr {\bf 0} & |B^{}_{l(\rm D)}| & A^{}_{l(\rm
D)} \cr} \right ) \; .
%       (5)
\end{equation}
We diagonalize $\overline{M}^{}_{l(D)}$ by using the orthogonal
transformation $O_l^{\dagger} \overline{M}^{}_l O_l^* = {\rm
Diag}\, \{m^{}_e, m^{}_{\mu}, m^{}_{\tau} \}$ or $O_{\rm
D}^{\dagger} \overline{M}^{}_{\rm D} O_{\rm D}^* = {\rm Diag}\,
\{d_1, d_2, d_3\}$, where $m^{}_{\alpha}$ (for $\alpha = e, \mu,
\tau$) and $d^{}_i$ (for $i = 1, 2, 3$) denote the masses of
charged leptons and the eigenvalues of $M^{}_{\rm D}$,
respectively. The elements of $O^{}_l$ can be given in terms of
two parameters $x^{}_l \equiv m^{}_e/m^{}_\mu$ and $y^{}_l \equiv
m^{}_\mu/m^{}_\tau$, and the elements of $O^{}_{\rm D}$ can
similarly be given in terms of two parameters $x \equiv
d^{}_1/d^{}_2$ and $y \equiv d^{}_2/d^{}_3$. Omitting the
subscripts of $O^{}_{l(\rm D)}$ and those of $(x^{}_l, y^{}_l)$,
we have \cite{Xing02}
\begin{eqnarray}
O^{}_{11} & = & + \left [ \frac{1-y}{(1+x)(1-xy)(1-y+xy)} \right
]^{1/2} \; ,
\nonumber \\
O^{}_{12} & = & -i \left [ \frac{x(1+xy)}{(1+x)(1+y)(1-y+xy)}
\right ]^{1/2} \; ,
\nonumber \\
O^{}_{13} & = & + \left [ \frac{xy^3 (1-x)}{(1-xy)(1+y)(1-y+xy)}
\right ]^{1/2} \; ,
\nonumber \\
O^{}_{21} & = & + \left [ \frac{x(1-y)}{(1+x)(1-xy)} \right
]^{1/2} \; ,
\nonumber \\
O^{}_{22} & = & +i \left [ \frac{1+xy}{(1+x)(1+y)} \right ]^{1/2}
\; ,
\nonumber \\
O^{}_{23} & = & + \left [ \frac{y(1-x)}{(1-xy)(1+y)} \right
]^{1/2} \; ,
\nonumber \\
O^{}_{31} & = & - \left [
\frac{xy(1-x)(1+xy)}{(1+x)(1-xy)(1-y+xy)} \right ]^{1/2} \; ,
\nonumber \\
O^{}_{32} & = & -i \left [ \frac{y(1-x)(1-y)}{(1+x)(1+y)(1-y+xy)}
\right ]^{1/2} \; ,
\nonumber \\
O^{}_{33} & = & + \left [ \frac{(1-y)(1+xy)}{(1-xy)(1+y)(1-y+xy)}
\right ]^{1/2} \; .
%       (6)
\end{eqnarray}
Note that $O^{}_{12}$, $O^{}_{22}$ and $O^{}_{32}$ are imaginary,
because the determinant of $\overline{M}^{}_{l(\rm D)}$ is
negative. The mass matrix $M^{}_{l(\rm D)}$ is therefore
diagonalized by the unitary matrix $V^{}_{l(\rm D)} = P^{}_{l(D)}
O_{l(\rm D)}$. In the FTY ansatz, $M^{}_{\rm R} = M^{}_0 {\bf 1}$
is taken and $M^{}_{\rm D}$ is assumed to be real (i.e.,
$P^{}_{\rm D} = {\bf 1}$ or $\phi^{}_{\rm D} = \varphi^{}_{\rm D}
= 0$, or equivalently $M^{}_{\rm D} = \overline{M}_{\rm D}$). Such
an assumption implies that one may diagonalize $M^{}_\nu$ in Eq.
(3) just by using the orthogonal transformation $V^{}_\nu =
O^{}_{\rm D} Q$ with $Q = {\rm Diag} \{ 1, i, 1 \}$:
\begin{equation}
V^\dagger_\nu M^{}_\nu V^*_\nu \; = \; \frac{\left [ Q^\dagger
\left (O^\dagger_{\rm D} \overline{M}^{}_{\rm D} O^*_{\rm D}
\right ) Q^* \right ]^2}{M^{}_0} \; =\; \frac{1}{M^{}_0}\left (
\matrix{ d^2_1 & 0 & 0 \cr 0 & d^2_2 & 0 \cr 0 & 0 & d^2_3 \cr}
\right ) \; ,
%       (7)
\end{equation}
from which the neutrino masses $m^{}_i = d^2_i/M^{}_0$ can be
obtained. The MNS matrix $V \equiv V^\dagger_l V^{}_\nu$ turns out
to be $V \equiv O^\dagger_l P^\dagger_l O^{}_{\rm D} Q$.

Next we go beyond the FTY ansatz by allowing the masses of three
heavy right-handed Majorana neutrinos (i.e., $M^{}_1$, $M^{}_2$
and $M^{}_3$) to be partially non-degenerate. For simplicity, we
keep the assumption of $M^{}_{\rm D} = \overline{M}^{}_{\rm D}$
(i.e., $M^{}_{\rm D}$ is real and $P^{}_{\rm D} = {\bf 1}$ holds)
and consider three different patterns of $M^{}_i$: (A) $M^{}_3 =
M^{}_2 \neq M^{}_1$, (B) $M^{}_2 = M^{}_1 \neq M^{}_3$ and (C)
$M^{}_1 = M^{}_3 \neq M^{}_2$. We also assume $M^{}_{\rm R}$ to be
diagonal, real and positive in the flavor basis chosen above.
Given $V_{\rm D}^{\dagger} M^{}_{\rm D} V_{\rm D}^* = {\rm Diag}
\{d^{}_1, d^{}_2, d^{}_3 \}$ with $V^{}_{\rm D} = P^{}_{\rm D}
O^{}_{\rm D} = O^{}_{\rm D}$ as we have already discussed, the
effective neutrino mass matrix $M^{}_\nu$ can now be expressed as
\begin{equation}
M^{}_\nu \; = \; M^{}_{\rm D} M^{-1}_{\rm R} M^T_{\rm D} \; = \;
\overline{M}^{}_{\rm D} M^{-1}_{\rm R} \overline{M}^T_{\rm D} \; =
\; O^{}_{\rm D} Q M'_\nu Q^T O^T_{\rm D} \; ,
%       (8)
\end{equation}
where $Q = {\rm Diag} \{ 1, i, 1\}$ and
\begin{equation}
M'_\nu \; \equiv \; Q^\dagger \left ( \matrix{ d^{}_1 & 0 & 0 \cr
0 & d^{}_2 & 0 \cr 0 & 0 & d^{}_3 \cr} \right ) O^T_{\rm D}
M^{-1}_{\rm R} O^{}_{\rm D} \left ( \matrix{ d^{}_1 & 0 & 0 \cr 0
& d^{}_2 & 0 \cr 0 & 0 & d^{}_3 \cr} \right ) Q^* \; .
%       (9)
\end{equation}
It is easy to check that Eq. (7) can simply be reproduced from
Eqs. (8) and (9) by taking $M^{}_{\rm R} = M^{}_0 {\bf 1}$. If the
masses of two right-handed Majorana neutrinos are not exactly
degenerate, their small difference will enter the expression of
$M'_\nu$. To be more transparent, we specify cases (A), (B) and
(C) as
\begin{eqnarray}
{\rm Case ~ (A):} ~~~~~~ M^{}_{\rm R} & = & \left ( \matrix{
M^{}_1 & 0 & 0 \cr 0 & M^{}_2 & 0 \cr 0 & 0 & M^{}_2 \cr} \right )
\; , \nonumber \\
{\rm Case ~ (B):} ~~~~~~ M^{}_{\rm R} & = & \left ( \matrix{
M^{}_1 & 0 & 0 \cr 0 & M^{}_1 & 0 \cr 0 & 0 & M^{}_2 \cr} \right )
\; , \nonumber \\
{\rm Case ~ (C):} ~~~~~~ M^{}_{\rm R} & = & \left ( \matrix{
M^{}_1 & 0 & 0 \cr 0 & M^{}_2 & 0 \cr 0 & 0 & M^{}_1 \cr} \right )
\; ,
%       (10)
\end{eqnarray}
and then define a single mass splitting parameter $\delta^{}_{12}
\equiv (M^{}_2 - M^{}_1)/M^{}_2$ for each case. We obtain
\begin{equation}
M'_\nu \; = \; \frac{d^2_3}{M^{}_1} \left ( \matrix{ x^2 y^2 & 0 &
0 \cr 0 & y^2 & 0 \cr 0 & 0 & 1 \cr} \right ) - ~ \delta^{}_{12}
M^{''}_\nu \; ~~~
%       (11)
\end{equation}
to the leading order of $\delta^{}_{12}$, and list the explicit
expressions of $M^{''}_\nu$ for three different cases in TABLE I.
$M^{'}_\nu$ can be diagonalized by the real orthogonal
transformation $O_{\nu}^{'T} M^{'}_\nu O_{\nu}^{'}={\rm Diag} \{
m_1, m_2, m_3\}$, where $m_i$ (for $i=1,2,3$) denote the neutrino
masses and $O_{\nu}^{'} = R^{}_{23}(\theta'_{23})
R^{}_{12}(\theta'_{12}) R^{}_{13}(\theta'_{13})$ is the product of
three canonical Euler rotation matrices. Thus it is the unitary
matrix $V^{}_\nu = O^{}_{\rm D} Q O'_\nu$ that diagonalizes the
neutrino mass matrix $M^{}_\nu$. The mass eigenvalues of
$M^{'}_\nu$ and the rotation angles of $O'_\nu$ are given in TABLE
II, in which the sub-leading terms of ${\cal O}(x^3)$, ${\cal O}(
y^3)$, ${\cal O}( x^2 y)$, ${\cal O}(xy^2)$ and ${\cal
O}(\delta^2_{12})$ have been neglected by assuming $0<x<1$ and
$0<y<1$ (i.e., $d_1<d_2<d_3$) as well as $|\delta_{12}| \ll 1$.
One may use the experimental values of $\Delta m^2_{21}$ and
$|\Delta m^2_{32}|$ to constrain the parameter space of $x$ and
$y$. Defining the ratio
\begin{equation}
R_{\nu} \; \equiv \; \frac{\Delta m_{21}^2}{|\Delta m_{32}^2|} \;
= \; y_{\nu}^2 \frac{1-x_{\nu}^2}{|1-y_{\nu}^2|} \; ,
%       (12)
\end{equation}
where $x^{}_{\nu} \equiv m_1/m_2$ and $y^{}_{\nu} \equiv m_2/m_3$,
we can numerically verify that only $y^{}_\nu < 1$ is allowed as a
consequence of $x <1$ and $y<1$.

Taking into account the mass splitting effect, we now express the
MNS matrix $V = V^\dagger_l V^{}_\nu$ as $V = O_l^{\dagger}
P_l^{\dagger} O^{}_{\rm D} Q O_{\nu}^{'}$, where $P^{}_l$ is taken
to be $P^{}_l = {\rm Diag} \{ e^{i\alpha}, e^{i\beta}, 1\}$ with
$\alpha = \phi^{}_l - \varphi^{}_l$ and $\beta = -\phi^{}_l$. $V$
totally contains seven parameters: $x^{}_l$, $y^{}_l$, $x$, $y$,
$\delta^{}_{12}$, $\alpha$ and $\beta$, among which $x^{}_l
\approx 0.00484$ and $y^{}_l \approx 0.0595$ \cite{PDG} are
already known to a high degree of accuracy. Given the value of
$\delta^{}_{12}$, one is able to constrain the ranges of $(x, y)$
and $(\alpha, \beta)$ by using current experimental data on
$R^{}_\nu$ and $(\theta^{}_{12}, \theta^{}_{23}, \theta^{}_{13})$.
To be specific, here we make use of the following parametrization
of $V$ \cite{Para}:
\begin{equation}
V = \left ( \matrix{c^{}_{12} c^{}_{13} & s^{}_{12} c^{}_{13} &
s^{}_{13} \cr -c^{}_{12} s^{}_{23} s^{}_{13} - s^{}_{12} c^{}_{23}
e^{-i\delta} & -s^{}_{12} s^{}_{23} s^{}_{13} + c^{}_{12}
c^{}_{23} e^{-i\delta} & s^{}_{23} c^{}_{13} \cr -c^{}_{12}
c^{}_{23} s^{}_{13} + s^{}_{12} s^{}_{23} e^{-i\delta} &
-s^{}_{12} c^{}_{23} s^{}_{13} - c^{}_{12} s^{}_{23} e^{-i\delta}
& c^{}_{23} c^{}_{13} \cr} \right ) \left ( \matrix{ e^{i\rho} & 0
& 0 \cr 0 & e^{i\sigma} & 0 \cr 0 & 0 & 1 \cr} \right ) \; ,
%       (13)
\end{equation}
where $c^{}_{ij} \equiv \cos \theta^{}_{ij}$, $s^{}_{ij} \equiv
\sin \theta^{}_{ij}$ (for $ij = 12, 23, 13$), $\delta$ denotes the
Dirac CP-violating phase, $\rho$ and $\sigma$ stand for the
Majorana CP-violating phases. Then
\begin{equation}
\sin^2 \theta^{}_{12} = \frac{|V^{}_{e2}|^2}{1-|V^{}_{e3}|^2} \; ,
~~~~ \sin^2 \theta^{}_{23} =
\frac{|V^{}_{\mu3}|^2}{1-|V^{}_{e3}|^2} \; , ~~~~ \sin^2
\theta^{}_{13} = |V^{}_{e3}|^2 \; .
%       (14)
\end{equation}
The leptonic Jarlskog invariant of CP violation \cite{J} reads
${\cal J} = s^{}_{12} c^{}_{12} s^{}_{23} c^{}_{23} s^{}_{13}
c^2_{13} \sin\delta$, which only depends on the Dirac phase
$\delta$. On the other hand, the effective masses of the tritium
beta decay $\langle m \rangle^{}_e$ and the neutrinoless
double-beta decay $\langle m \rangle^{}_{ee}$ are given by
\begin{equation}
\langle m \rangle_e^2 = \sum_{i=1}^3 \left (m_i^2 |V_{ei}|^2
\right ) = m_3^2 \left (x_{\nu}^2 y_{\nu}^2 c^2_{12} c^2_{13} +
y_{\nu}^2 s^2_{12} c^2_{13} + s^2_{13} \right ) \; , ~~~~~~
%       (15)
\end{equation}
and
\begin{equation}
\langle m \rangle^{}_{ee} = \Biggl| \sum_{i=1}^3 \left (m^{}_i
V_{ei}^2 \right ) \Biggr| = m^{}_3 \left | x^{}_{\nu} y^{}_{\nu}
c^2_{12} c^2_{13} e^{2i\rho} + y^{}_{\nu} s^2_{12} c^2_{13}
e^{2i\sigma} + s^2_{13} \right | \; .
%       (16)
\end{equation}
One can see that the Dirac phase $\delta$ does not appear in
$\langle m\rangle^{}_{ee}$, as guaranteed by the phase convention
of $V$ in Eq. (13). The present experimental upper bounds on
$\langle m \rangle^{}_e$ and $\langle m \rangle^{}_{ee}$ are
$\langle m \rangle^{}_e < 2.1~{\rm eV}$ and $\langle m
\rangle^{}_{ee} < 0.39~{\rm eV}$, respectively, at the $99 \%$
confidence level \cite{vissani}. The future KATRIN experiment is
expected to reach the sensitivity $\langle m \rangle^{}_e \simeq
0.2~{\rm eV}$ \cite{KATRIN}. On the other hand, the new
experiments towards searching for the neutrinoless double-beta
decay may reach the sensitivity $\langle m \rangle^{}_{ee} \sim
{\cal O} (10^{-2})~{\rm eV}$ \cite{0nubeta}.

\section{Numerical analysis}

We proceed to do a numerical analysis of the generalized FTY
ansatz with non-vanishing $\delta^{}_{12}$. The neutrino mass
spectrum and the MNS matrix depend on seven free parameters:
$d^{}_3$, $x$, $y$, $\alpha$, $\beta$, $M_1$ and $\delta^{}_{12}$.
These parameters may more or less get constrained from the
experimental data given in Eqs. (1) and (2). In the
$\delta^{}_{12} = 0$ case, which corresponds to the original FTY
ansatz, an updated numerical analysis yields $\omega \equiv
d_3^2/M_1(=m^{}_3)\approx (47 \sim 56) ~{\rm meV}$,
$x\,(=\sqrt{x^{}_{\nu}})\approx 0.22 \sim 0.56$,
$y\,(=\sqrt{y^{}_{\nu}})\approx 0.39 \sim 0.45$, $\alpha \approx 0
\sim 2\pi$ and $\beta \approx 0.61\pi \sim 1.4\pi$. These results
allow us to make the following predictions:
\begin{eqnarray}
&& \sin^2 \theta^{}_{23} \approx 0.35 \sim 0.49 \; , \quad \sin^2
\theta^{}_{12} \approx 0.25 \sim 0.38 \; , \quad \sin^2
\theta^{}_{13} \approx 0.0050 \sim 0.030 \; ;
\nonumber \\
&& R^{}_{\nu} \approx 0.023 \sim 0.041 \; , \quad x^{}_{\nu}
\approx 0.049 \sim 0.32 \; , \quad y^{}_{\nu} \approx 0.15 \sim
0.20 \; ;
\nonumber \\
&& \delta \approx -0.26\pi \sim 0.26\pi \; , \quad \rho \approx
-0.12\pi \sim 0.12\pi \; , \quad
\sigma \approx -0.16\pi \sim 0.16\pi \; ;
\nonumber \\
&& \langle m \rangle^{}_e \approx (5.8 \sim 11) ~{\rm meV} \; ,
\quad \langle m \rangle^{}_{ee} \approx (2.8 \sim 6.6) ~{\rm meV}
\; , \quad {\cal J} \approx (-2.3 \sim 2.3) \times 10^{-2} \; .
%       (17)
\end{eqnarray}
Note that the absolute scales of $d^{}_3$ and $M^{}_1$ cannot
separately be determined from our analysis. FIG. 1 shows the
allowed regions of $\sin^2 \theta_{12}$ vs $\sin^2 \theta_{23}$,
$R_{\nu}$ vs $\sin^2 \theta_{13}$ and $\delta$ vs $\langle m
\rangle_{ee}$ in the $\delta^{}_{12} = 0$ case. To get a ball-park
feeling of the effect of non-vanishing $\delta^{}_{12}$ on
neutrino masses and lepton flavor mixing, we may carry out a
general numerical calculation by allowing $\delta^{}_{12}$ to vary
between $-1$ and $+1$. Then we obtain the generous constraints on
$\delta^{}_{12}$ and $\omega$ as follows:
\begin{eqnarray}
& & {\rm Case ~ (A)}: ~~ \delta_{12} \leq 0.41 \; , \quad \omega
\leq 8.7 \times 10^{-2}~{\rm eV} \; ;
\nonumber \\
& & {\rm Case ~ (B)}: ~~ -0.82 \leq \delta_{12} \leq 0.99 \; ,
\quad 3.2 \times 10^{-2}~{\rm eV} \leq \omega \leq 2.2 \times
10^{-1}~{\rm eV} \; ;
\nonumber \\
& & {\rm Case ~ (C)}: ~~ -0.81 \leq \delta_{12} \leq 0.52 \; ,
\quad 4.1 \times 10^{-2}~{\rm eV} \leq \omega \leq 6.3 \times
10^{-2}~{\rm eV} \; .
%       (18)
\end{eqnarray}
We have found that the parameter space in the $\delta^{}_{12} \neq
0$ case is essentially distinguishable from that in the
$\delta^{}_{12} = 0$ case, provided $|\delta^{}_{12}|$ is of
${\cal O}(0.1)$ or larger. However, it seems more interesting to
consider a small mass splitting between two heavy right-handed
Majorana neutrinos in model building; i.e., $|\delta^{}_{12}| \sim
{\cal O}(0.1)$. Thus we shall fix $|\delta^{}_{12}| =0.25$ as a
typical input in our subsequent numerical calculations. Such a
choice of $|\delta^{}_{12}|$ implies that the analytical
approximations made in Eq. (11) and TABLE II are valid.

TABLE III shows the allowed ranges of model parameters in cases
(A), (B) and (C) with $|\delta^{}_{12}|=0.25$. Note that cases
(A$^{\pm}$), (B$^{\pm}$) or (C$^{\pm}$) correspond to the positive
and negative values of $\delta^{}_{12}$ in this table, where the
predictions for a number of observables are also listed. Some
comments are in order.
\begin{itemize}
\item       The model parameters $\omega$ and $y$ are restricted
to the relatively narrow ranges. The allowed region of $x$ is
somehow larger than that of $y$ in all cases, implying that the
flavor mixing angle $\theta^{}_{12}$ is less constrained than
$\theta^{}_{23}$ in the generalized FTY ansatz. There is little
limitation to the phase parameter $\alpha$, although $\alpha =
\pi$ has been ruled out in cases (A$^\pm$), (B$^-$) and (C$^-$).
In comparison, the phase parameter $\beta$ always takes values
around $\pi$ to guarantee $\sin^2 \theta^{}_{23} > 0.35$.

\item       The maximal atmospheric neutrino mixing (i.e.,
$\theta^{}_{23} = \pi/4$ or $\sin^2\theta^{}_{23} = 0.5$) can only
be achieved in cases (B$^+$) and (C$^+$). In all cases,
$\sin^2\theta^{}_{12}$ is not well restricted. But the smallest
neutrino mixing angle $\theta^{}_{13}$ has an lower bound in each
case: $\theta^{}_{13} > 3.76^\circ$ (A$^+$), $3.85^\circ$ (A$^-$),
$2.43^\circ$ (B$^+$), $5.13^\circ$ (B$^-$), $2.81^\circ$ (C$^+$)
and $4.66^\circ$ (C$^-$). If a reactor neutrino oscillation
experiment can reach the sensitivity $\sin^2 2\theta^{}_{13} \sim
1\%$ (or equivalently $\theta^{}_{13} \sim 2.87^\circ$), it will
be able to test our prediction for $\theta^{}_{13}$.

\item       The ranges of three CP-violating phases are not very
restrictive. In particular, the CP-conserving case (i.e., $\delta
= \rho = \sigma =0$) cannot be ruled out at present. The allowed
range of ${\cal J}$ is roughly the same in all six cases. It is in
principle possible to detect $|{\cal J}| \sim {\cal O}(10^{-2})$
in the future long-baseline neutrino oscillation experiments.

\item       Just as we have expected, the neutrino mass spectrum
has a normal hierarchy with $m^{}_3 \approx \sqrt{|\Delta
m^2_{32}|} ~\sim 0.05$ eV. The magnitude of $\langle
m\rangle^{}_e$ is therefore strongly suppressed, at most of ${\cal
O}(10^{-2})$ eV, which cannot be probed by the proposed KATRIN
experiment. In comparison, the magnitude of $\langle
m\rangle^{}_{ee}$ is at the level of a few meV, which might be
measured in the future neutrinoless double-beta decay experiments.
\end{itemize}

To see how the parameter space changes from the $\delta^{}_{12} =
0$ case to the $\delta^{}_{12} = \pm 0.25$ cases, we plot the
allowed regions of $\sin^2 \theta^{}_{12}$ vs $\sin^2
\theta^{}_{23}$, $R^{}_{\nu}$ vs $\sin^2 \theta^{}_{13}$ and
$\delta$ vs $\langle m \rangle^{}_{ee}$ in FIGs. 1--4. The points
in the scatter plots for different parameters may more or less
correlate, because they all depend on the model parameters
$\delta^{}_{12}$, $\omega$, $x$, $y$, $\alpha$ and $\beta$. One
can easily see from TABLE II that the magnitude of $\theta'_{23}$
is suppressed by the smallness of $xy$ in case (A), hence the
enhancement or suppression effect of $\delta^{}_{12} \neq 0$ on
$\theta^{}_{23}$ is not obvious at all in FIG. 2. Similar
arguments hold for $\theta^{}_{12}$ in case (B) and
$\theta^{}_{13}$ in case (A). Note that the deviation of $m^{}_3$
from $\omega$ in case (C) is also suppressed by the smallness of
$y$. These analytical features are consistent with the numerical
results shown in FIGs. 2--4. We find that case (C) in FIG. 4 is
most sensitive to the effect induced by $\delta^{}_{12} \neq 0$.

It is worth mentioning that our results, similar to those obtained
from the FTY ansatz, are essentially stable against radiative
corrections from the seesaw scale to the electroweak scale or vice
versa. The reason is simply that $m^{}_1$, $m^{}_2$ and $m^{}_3$
have a clear normal hierarchy (or equivalently, $x^{}_\nu < 1$ and
$y^{}_\nu < 1$ in TABLE III) \cite{RGE}. Therefore, we have
omitted the insignificant renormalization-group running effects on
three light neutrino masses, three flavor mixing angles and three
CP-violating phases in our calculations. Note that the structural
hierarchy of $M^{}_\nu$ is not strong, nor is that of $M^{}_{\rm
D}$. To see this point more clearly, we have evaluated the ratios
of $A^{}_{\rm D}$, $|B^{}_{\rm D}|$ and $|C^{}_{\rm D}|$ to
$d^{}_3$ and listed them in TABLE III. Taking case (B$^+$) for
example, we have
%%%%%%%%%%%%%%%
\begin{eqnarray}
M^{}_{\rm D} \; \sim \; d^{}_3 \left ( \matrix{ {\bf 0} & 0.21
\cdots 0.36 & {\bf 0} \cr 0.21 \cdots 0.36 & {\bf 0} & 0.34 \cdots
0.55 \cr {\bf 0} & 0.34 \cdots 0.55 & 0.68 \cdots 0.87 \cr}
\right) \; ,
%       (19)
\end{eqnarray}
where the absolute value of $d^{}_3$ is unrestricted (note that
only $d^2_3/M^{}_1 \equiv \omega$ gets constrained from current
experimental data, as shown in TABLE III). Comparing $M^{}_{\rm
D}$ with the Fritzsch-type up-quark mass matrix \cite{F78}
\begin{eqnarray}
M^{}_{\rm up} \; \sim \; m^{}_t \left ( \matrix{ {\bf 0} & ~~ 2.0
\times 10^{-4} ~~ & {\bf 0} \cr 2.0 \times 10^{-4} & {\bf 0} & 6.5
\times 10^{-2} \cr {\bf 0} & 6.5 \times 10^{-2} & 1 \cr} \right)
\; ,
%       (20)
\end{eqnarray}
where $m^{}_u/m^{}_c \sim 0.0023$ and $m^{}_c/m^{}_t \sim 0.0042$
are typically input, one can observe that the hierarchy of
$M^{}_{\rm D}$ is much weaker than that of $M^{}_{\rm up}$. As
$M^{}_l$ is strongly hierarchical \cite{Xing02} and $M^{}_{\rm R}$
is close to the unity matrix in our scenario, the weak hierarchy
of $M^{}_{\rm D}$ is the main source of large flavor mixing in the
lepton sector.

Finally, we stress that our work will make much more sense, if
some tension or disagreement appears between the original FTY
ansatz and more accurate neutrino oscillation data in the near
future. One may also generalize the FTY ansatz by abandoning the
exact mass degeneracy of three right-handed Majorana neutrinos and
allowing for $M^{}_1 \sim M^{}_2 \sim M^{}_3$. This case, which
involves two small mass splitting parameters, can be regarded as a
further extension of cases (A), (B) and (C) discussed in the
present paper.

\section{Summary}

We have generalized the interesting FTY ansatz by allowing the
masses of three heavy right-handed Majorana neutrinos to be
partially non-degenerate, and then investigated the
phenomenological consequences of this mass splitting on the
neutrino mass spectrum, flavor mixing angles and CP-violating
phases. Three simple but typical cases have been considered in our
analysis: (A) $M^{}_3 = M^{}_2 \neq M^{}_1$, (B) $M^{}_2 = M^{}_1
\neq M^{}_3$ and (C) $M^{}_1 = M^{}_3 \neq M^{}_2$. The analytical
approximations and numerical results show that there is the
parameter space in every case and the mass splitting effect may
play an important role in fitting the experimental data. We have
also obtained the numerical predictions for the Jarlskog invariant
of CP violation and the effective masses of the tritium beta decay
and the neutrinoless double-beta decay. Some of our results can be
experimentally tested in the near future.

In our discussions, we have taken both the Dirac neutrino mass
matrix $M^{}_{\rm D}$ and the right-handed Majorana neutrino mass
matrix $M^{}_{\rm R}$ to be real. Hence there is no leptogenesis
even though the mass degeneracy of three right-handed neutrinos is
partially broken. It is of course interesting to introduce
non-trivial complex phases into $M^{}_{\rm D}$ and (or) $M^{}_{\rm
R}$, such that CP violation may appear in the
lepton-number-violating and out-of-equilibrium decays of heavy
right-handed Majorana neutrinos. We shall elaborate this idea
elsewhere, so as to accommodate baryogenesis via leptogenesis in a
new extension of the FTY ansatz.

\begin{acknowledgments}

We would like to thank S. Zhou for useful discussions. This work
is supported in part by the National Natural Science Foundation of
China.

\end{acknowledgments}

\newpage
%%%%%%%%%%%%%%%%%%%%%%% TABLE I %%%%%%%%%%%%%%%%%%%%%%%%%%%%%%%%%%%%%%%%%
%%%    Table for M'_\nu in Cases A, B, C
%%%%%%%%%%%%%%%%%%%%%%%%%%%%%%%%%%%%%%%%%%%%%%%%%%%%%%%%%%%%%%%%
\begin{table}
\caption{The expressions of $M_{\nu}^{''}$ for cases (A), (B) and
(C). Here we have defined $\omega \equiv d_3^2/M_1$, $D^{}_{\nu}
\equiv (1+x)(1-xy)(1+y)(1-y+xy)$, $D_{\nu}' \equiv
(1+x)(1-xy)(1+y)$, $F^{}_{x,y} \equiv \sqrt{xy(1-x^2)(1-y^2)}~$,
$F^{}_{x,xy} \equiv \sqrt{y(1-x^2)(1-x^2y^2)}~$ and $F^{}_{y,xy}
\equiv \sqrt{x(1-y^2)(1-x^2y^2)}~$, where $x=d_1/d_2$ and
$y=d_2/d_3$.} \vspace{0.3cm}
\setlength{\tabcolsep}{2pt}\footnotesize
\begin{tabular}{|c|l|l|}
Case & ~~~~~~~~~~~~~~~~~~~~~~~~~~~~~~~~~~~
${M^{''}_{\nu}}^{\mathstrut}_{\mathstrut}$ & ~~~~~ Abbreviated
functions
\\ \hline
%--------------------------------------------------------------
(A) & $\displaystyle \frac{\omega}{D^{}_{\nu}} \left ( \matrix{
x^3 y^2 F_{11}^{\rm A} & xy^2 F^{}_{y, xy} & -xy^2 F^{}_{x, y} \cr
xy^2 F^{}_{y, xy} & y^2 F_{22}^{\rm A} & x y^2 F^{}_{x,xy} \cr
-xy^2 F^{}_{x, y} & x y^2 F^{}_{x,xy} & F_{33}^{\rm A} \cr} \right
)^{\mathstrut}_{\mathstrut}$ &
$\begin{array}{l} F_{11}^{\rm A} \equiv (1+y) (1-y-x^2y^2+y^2 ) \\
F_{22}^{\rm A}\equiv (1-xy) (1+xy+x^2 y^2-y^2 ) \\ F_{33}^{\rm A}
\equiv (1+x) (1-y^2+xy^2-x^2y^2 ) \end{array}$
\\ \hline
%%%---------
(B) & $\displaystyle\frac{\omega}{D^{}_{\nu}} \left ( \matrix{ x^3
y^3 F_{11}^{\rm B} & xy^3(1-x) F^{}_{y,xy} & -xy(1+xy) F^{}_{x,y}
\cr xy^3(1-x) F^{}_{y,xy} & y^3 F_{22}^{\rm B} & y(1-y) F_{x,xy}
\cr -xy(1+xy) F^{}_{x,y} & y(1-y) F^{}_{x,xy} & F_{33}^{\rm B} }
\right )^{\mathstrut}_{\mathstrut}$ & $\begin{array}{l}
F_{11}^{\rm B} \equiv (1-x)(1+xy)(1+y) \\ F_{22}^{\rm B} \equiv
(1-x)(1-xy)(1-y) \\ F_{33}^{\rm B} \equiv (1+x)(1+xy)(1-y)
\end{array}$
\\
\hline
%%%---------
(C) & $ \displaystyle\frac{\omega}{D'_{\nu}} \left ( \matrix{
x^3y^2 F_{11}^{\rm C} & xy^2 F^{}_{y,xy} & xy F^{}_{x,y}  \cr xy^2
F^{}_{y,xy} & y^2 F_{22}^{\rm C} & y F^{}_{x,xy} \cr xy F^{}_{x,y}
& y F^{}_{x,xy} & F_{33}^{\rm C} } \right
)^{\mathstrut}_{\mathstrut}$ & $\begin{array}{l} F_{11}^{\rm C}
\equiv (1-y^2) \\ F_{22}^{\rm C} \equiv (1-x^2y^2) \\ F_{33}^{\rm
C} \equiv y(1-x^2) \end{array}$
\\
\end{tabular}
\end{table}

%%%%%%%%%%%%%%%%%%%%%%%%%% TABLE II %%%%%%%%%%%%%%%%%%%%%%%%%%%%%%%%%%%%%%
%%%    Table for m_1, m_2, m_3 and mixing angles in A, B, C
%%%%%%%%%%%%%%%%%%%%%%%%%%%%%%%%%%%%%%%%%%%%%%%%%%%%%%%%%%%%%%%%
\begin{table}
\caption{The mass eigenvalues and rotation angles of $M_{\nu}^{'}$
in cases (A), (B) and (C). Here the definitions of $\omega$,
$F^{}_{x,y}$, $F^{}_{x,xy}$ and $F^{}_{y,xy}$ are the same as
those in TABLE I. We have neglected the sub-leading terms of
${\cal O}(x^3)$, ${\cal O}(y^3)$, ${\cal O}(x^2 y)$, ${\cal
O}(xy^2)$ and ${\cal O}(\delta^2_{12})$ in our analytical
approximations.} \vspace{0.3cm}
\setlength{\tabcolsep}{4pt}\footnotesize
\begin{tabular}{|c|l|l|}
Case & Mass eigenvalues of $M_{\nu}^{'}$  & Rotation angles of
$M_{\nu}^{'}$
\\ \hline
& $m^{}_1 \simeq \omega x^2y^2 \biggl [ 1 - \frac{x(1-y+y^2) \,
\delta^{}_{12}} {(1+x)(1-xy)(1-y+xy)} \biggr
]^{\mathstrut}_{\mathstrut}$
& $\tan 2\theta'_{23} \simeq
\frac{-2x_{\mathstrut} y^2 F^{}_{x,xy} \, \delta^{}_{12} }
{(1+x-2y^2)-(1+x-2y^2) \, \delta_{12}^{\mathstrut}}$ \\
(A) & $m^{}_2 \simeq \omega y^2 \biggl [ 1 - \frac{(1+xy-y^2) \,
\delta^{}_{12}} {(1+x)(1+y)(1-y+xy)} \biggr
]^{\mathstrut}_{\mathstrut}$
& $\tan 2\theta'_{12} \simeq
\frac{-2x_{\mathstrut} F^{}_{y,xy} \, \delta^{}_{12} }
{ (1+x-x^2-y^2)+(1-y^2) \, \delta_{12}^{\mathstrut}}$ \\
 & $m^{}_3 \simeq \omega
\biggl [ 1 - \frac{(1-y^2) \, \delta^{}_{12}}
{(1-xy)(1+y)(1-y+xy)} \biggr ]^{\mathstrut}_{\mathstrut}$ & $\tan
2\theta'_{13} \simeq \frac{2x_{\mathstrut} y^2 F^{}_{x,y} \,
\delta^{}_{12} } {(1+x-y^2)-(1+x-y^2) \,
\delta_{12}^{\mathstrut}}$
\\
\hline
%%%---------
 &
$m^{}_1 \simeq \omega x^2y^2 \biggl [ 1- \frac{xy_{\mathstrut}
(1-x+xy) \, \delta^{}_{12}} {(1+x) (1-xy) (1-y+xy)} \biggr
]^{\mathstrut}_{\mathstrut}$ ~~~~
& $\tan 2\theta'_{23} \simeq
\frac{-2y_{\mathstrut} (1-y) F^{}_{x,xy} \, \delta^{}_{12} }
{(1+x-2y^2)-(1+x-y) \, \delta_{12}^{\mathstrut}}$
\\
(B) & $m^{}_2 \simeq  \omega y^2 \biggl [ 1-\frac{y (1-x)(1-y) \,
\delta_{12}} {(1+x) (1+y) (1-y+xy)} \biggr
]_{\mathstrut}^{\mathstrut}$ & $\tan 2\theta'_{12} \simeq \frac{
-2xy_{\mathstrut} (1-x) F^{}_{y,xy} \, \delta^{}_{12} }
{(1+x-x^2-y^2)-y(1-x-y) \, \delta_{12}^{\mathstrut}}$ ~~~~~
\\
& $m^{}_3 \simeq \omega \biggl [ 1-\frac{(1-y+xy) \,
\delta^{}_{12}} {(1-xy)(1+y)(1-y+xy)} \biggr
]^{\mathstrut}_{\mathstrut}$ & $\tan 2\theta'_{13} \simeq
\frac{2xy_{\mathstrut} (1+xy) F^{}_{x,y} \, \delta_{12} }
{(1+x-y^2)-(1+x-y) \, \delta_{12}^{\mathstrut}}$
\\
\hline
%%%---------
 &
$m^{}_1 \simeq \omega x^2y^2 \biggl [ 1-\frac{x(1-y) \,
\delta_{12}}{(1+x)(1-xy)} \biggr ]^{\mathstrut}_{\mathstrut}$ &
$\tan 2\theta'_{23} \simeq \frac{ -2y_{\mathstrut} F^{}_{x,xy} \,
\delta^{}_{12} } {(1+x+y-y^2)-y(1-y-x^2) \,
\delta_{12}^{\mathstrut}}$
\\
(C) & $m^{}_2 \simeq \omega y^2 \biggl [1-\frac{(1+xy) \,
\delta^{}_{12}} {(1+x)(1+y)}  \biggr ]_{\mathstrut}^{\mathstrut}$
& $\tan 2\theta'_{12} \simeq \frac{ -2x_{\mathstrut} F^{}_{y,xy}
\, \delta^{}_{12} } { (1+x+y-x^2)- \delta_{12}^{\mathstrut}}$
\\
 &
$m^{}_3 \simeq \omega  \biggl [ 1-\frac{y(1-x) \,
\delta^{}_{12}}{(1+y)(1-xy)}  \biggr ]^{\mathstrut}_{\mathstrut}$
& $\tan 2\theta'_{13} \simeq \frac{-2xy_{\mathstrut} F^{}_{x,y} \,
\delta^{}_{12} } {(1+x+y)-y(1-x^2) \, \delta_{12}^{\mathstrut}}$
\\
\end{tabular}
\end{table}

\newpage
%%%%%%%%%%%%%%%%%%%%%%%%%% TABLE III %%%%%%%%%%%%%%%%%%%%%%%%%%%%%%%%%%%%%%
%%%    Numerical results in A, B, C
%%%%%%%%%%%%%%%%%%%%%%%%%%%%%%%%%%%%%%%%%%%%%%%%%%%%%%%%%%%%%%%%
\begin{table}
\caption{ The allowed ranges of free parameters and the predicted
values of some observables in cases (A), (B) and (C) with the
typical input $|\delta^{}_{12}| = 0.25$. Here cases (A$^{\pm}$),
(B$^{\pm}$) or (C$^{\pm}$) correspond to the positive and negative
values of $\delta^{}_{12}$.} \vspace{0.3cm}
\setlength{\tabcolsep}{1pt}\footnotesize
\begin{tabular}{|c|c|c|c|c|c|c|}
%& \multicolumn{2}{c}{A} & \multicolumn{2}{c}{B} & \multicolumn{2}{c}{C} \\
 & (A$^+$) & (A$^-$) & (B$^+$) & (B$^-$) & (C$^+$) & (C$^-$) \\
\hline
%%%-------
$\delta^{}_{12}$ & $0.25$  & $-0.25$  & $0.25$
& $-0.25$  & $0.25$  & $-0.25$  \\
%%%-------
$\omega~({\rm meV})$ & $62 \sim 75$ & $40 \sim 45$ &
$57 \sim 72$ & $40 \sim 48$ & $49 \sim 60$ & $44 \sim 54$ \\
%%%-------
$x$ & $0.26 \sim 0.68$ & $0.20 \sim 0.47$ &
$0.23 \sim 0.67$ & $0.21 \sim 0.51$ & $0.22 \sim 0.75$ & $0.21 \sim 0.45$ \\
%%%-------
$y$ & $0.37 \sim 0.43$ & $0.40 \sim 0.44$ &
$0.36 \sim 0.41$ & $0.42 \sim 0.48$ & $0.42 \sim 0.51$ & $0.38 \sim 0.43$ \\
%%%-------
$\alpha$ ($\pi$) & $\left \{ \begin{array}{@{\,}c@{\,}} 0 \sim 0.9
\cr 1.1 \sim 2.0
\end{array} \right .$
& $\left \{ \begin{array}{@{\,}c@{\,}} 0 \sim 0.9 \cr 1.1 \sim 2.0
\end{array} \right .$
& $0 \sim 2.0$ & $\left \{ \begin{array}{@{\,}c@{\,}} 0.013 \sim
0.70 \cr 1.3 \sim 2.0
\end{array} \right .$ & $0 \sim 2.0$
& $\left \{ \begin{array}{@{\,}c@{\,}} 0.019 \sim 0.85 \cr 1.3
\sim 2.0
\end{array} \right .$ \\
%%%-------
$\beta$ ($\pi$) & $0.64 \sim 1.4$ & $0.63 \sim 1.4$
& $0.61 \sim 1.4$ & $0.67 \sim 1.4$ & $0.51 \sim 1.5$ & $0.69 \sim 1.3$ \\
\hline \hline
%%%-------
$\sin^2 \theta^{}_{23}$ & $0.35 \sim 0.49$ & $0.35 \sim 0.48$ &
$0.35 \sim 0.54$ & $0.35 \sim 0.47$ & $0.35 \sim 0.55$ & $0.35 \sim 0.44$ \\
%%%-------
$\sin^2 \theta^{}_{12}$ & $0.25 \sim 0.34$ & $0.25 \sim 0.38$ &
$0.25 \sim 0.38$ & $0.25 \sim 0.38$ & $0.25 \sim 0.36$ & $0.25 \sim 0.38$ \\
%%%-------
$\sin^2 \theta^{}_{13}$ & $0.0043 \sim 0.030$ & $0.0045 \sim
0.030$ &
$0.0018 \sim 0.030$ & $0.0080 \sim 0.030$ & $0.0024 \sim 0.030$ & $0.0066 \sim 0.030$ \\
%%%-------
$R^{}_{\nu}$ & $0.023 \sim 0.042$ & $0.024 \sim 0.036$ &
$0.024 \sim 0.042$ & $0.024 \sim 0.040$ & $0.024 \sim 0.042$ & $0.024 \sim 0.042$ \\
%%%-------
$m^{}_3~({\rm meV})$ & $47 \sim 57$ & $50 \sim 56$ &
$47 \sim 56$ & $47 \sim 56$ & $47 \sim 56$ & $47 \sim 56$ \\
%%%-------
$x^{}_{\nu}$ & $0.076 \sim 0.45$ & $0.033 \sim 0.20$ &
$0.052 \sim 0.43$ & $0.044 \sim 0.26$ & $0.058 \sim 0.56$ & $0.039 \sim 0.18$ \\
%%%-------
$y^{}_{\nu}$ & $0.15 \sim 0.22$ & $0.15 \sim 0.19$ &
$0.15 \sim 0.21$ & $0.15 \sim 0.20$ & $0.15 \sim 0.24$ & $0.15 \sim 0.20$ \\
%%%-------
$\delta$ ($\pi$) & $-0.26 \sim 0.25$ & $-0.26 \sim 0.25$ &
$-0.30 \sim 0.30$ & $-0.23 \sim 0.23$ & $-0.29 \sim 0.29$ & $-0.24 \sim 0.23$ \\
%%%-------
$\rho$ ($\pi$) & $-0.13 \sim 0.13$ & $-0.12 \sim 0.12$ & $-0.18
\sim 0.18$ & $-0.10 \sim 0.10$ & $-0.16 \sim 0.16$ &
$-0.09 \sim 0.09$ \\
%%%-------
$\sigma$ ($\pi$) & $-0.16 \sim 0.16$ & $-0.15 \sim 0.15$ & $-0.20
\sim 0.20$ & $-0.13 \sim 0.13$ &
$-0.19 \sim 0.18$ & $-0.13 \sim 0.13$ \\
%%%-------
$\langle m \rangle^{}_e \, ({\rm meV})$ & $5.5 \sim 10$ & $5.7
\sim 11$ & $4.9 \sim 11$ & $6.5 \sim 11$
& $5.2 \sim 11$ & $6.3 \sim 11$ \\
%%%-------
$\langle m \rangle^{}_{ee} \, ({\rm meV})$ & $2.9 \sim 6.7$ & $2.7
\sim 5.8$ & $2.7 \sim 7.6$ & $2.9 \sim 6.1$
& $2.7 \sim 8.0$ & $2.9 \sim 5.5$ \\
%%%-------
${\cal J} ~(10^{-2})$ & $-2.1 \sim 2.2$ & $-2.1 \sim 2.3$ & $-2.1
\sim 2.1$ & $-2.3 \sim 2.2$ & $-2.2 \sim 2.1$ & $-2.1 \sim 2.1$ \\
\hline \hline $A^{}_{\rm D}/d^{}_3$ & $0.68 \sim 0.86$ & $0.65
\sim 0.78$ & $0.68 \sim 0.87$
& $0.64 \sim 0.78$ & $0.63 \sim 0.87$ & $0.68 \sim 0.78$ \\
%%%-------
$|B^{}_{\rm D}|/d^{}_3$ & $0.34 \sim 0.55$ & $0.45 \sim 0.58$ &
$0.34 \sim 0.55$
& $0.43 \sim 0.57$ & $0.32 \sim 0.58$ & $0.45 \sim 0.55$ \\
%%%-------
$|C^{}_{\rm D}|/d^{}_3$ & $0.24 \sim 0.38$ & $0.22 \sim 0.33$ &
$0.21 \sim 0.36$
& $0.24 \sim 0.36$ & $0.25 \sim 0.47$ & $0.21 \sim 0.32$ \\
\end{tabular}
\end{table}
%%%%%%%%%%%%%%%%%

\newpage
%%%%%%%%%%%%%%%%%%%
% Fig. 1
%%%%%%%%%%%%%%%%%%%
\begin{figure}
\begin{center}
\includegraphics[width=6.5cm,clip]{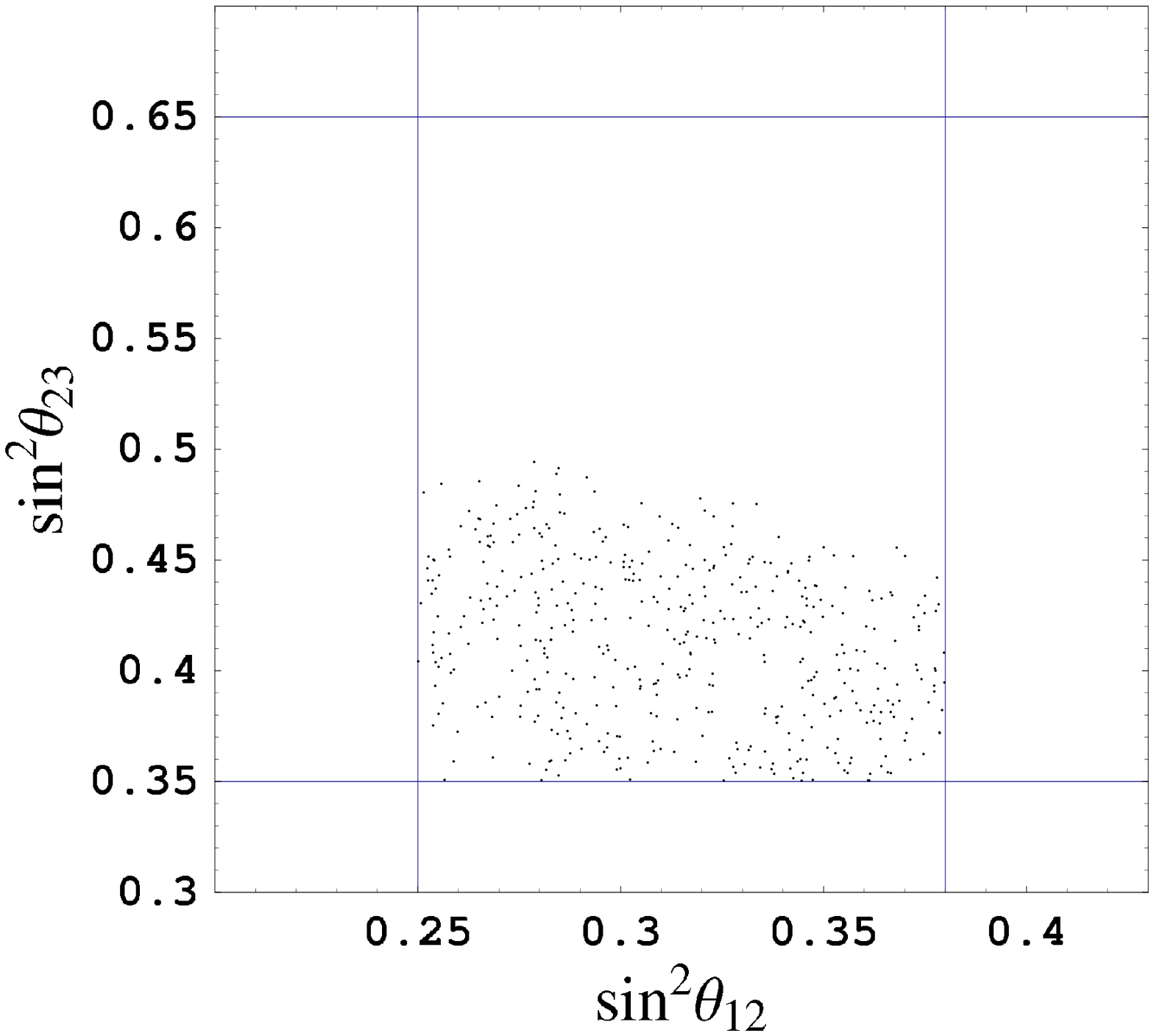}
\\
\vspace{0.4cm}
\hspace*{-1.5mm} %
\includegraphics[width=6.5cm,clip]{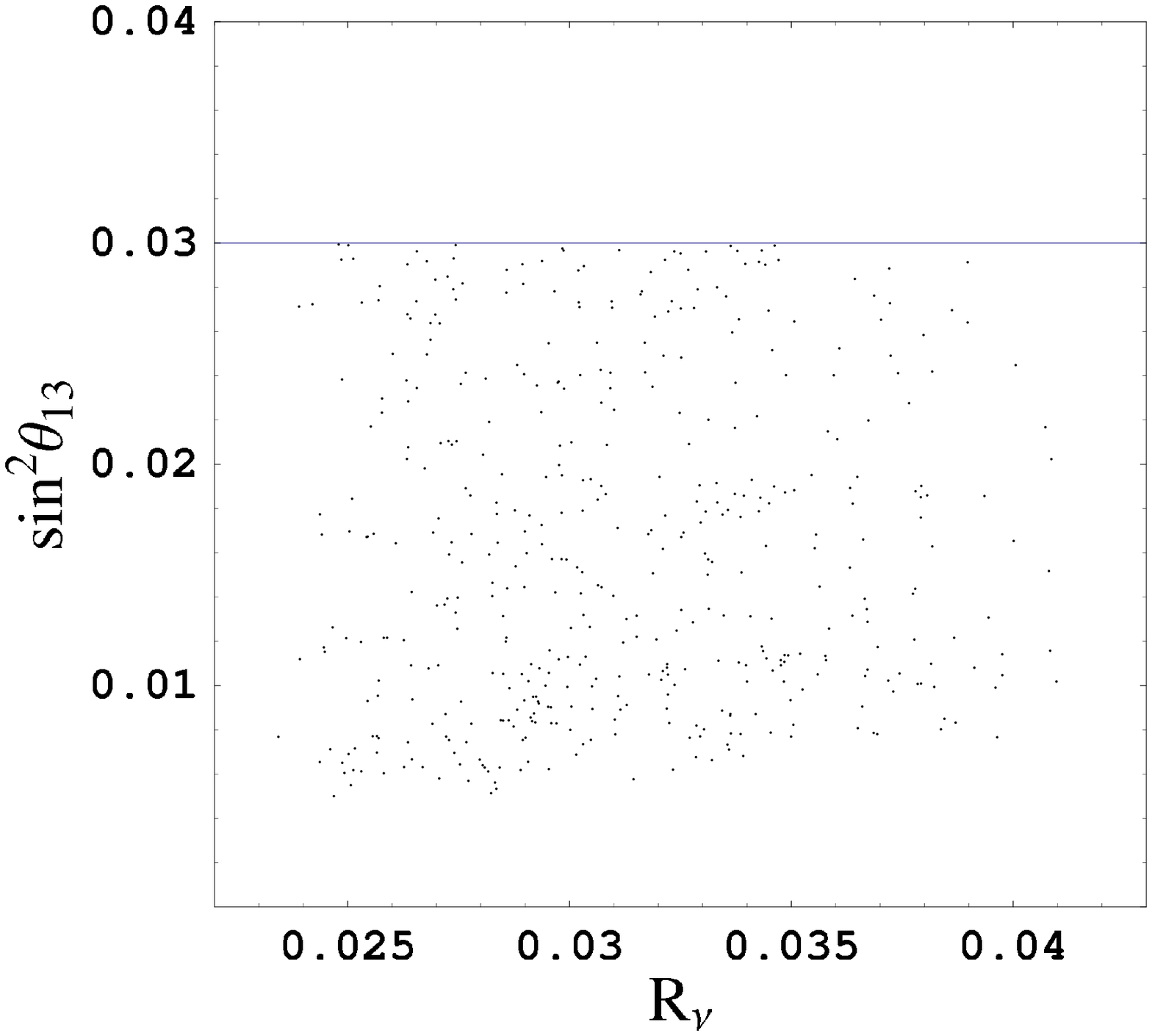}
\\
\vspace{0.4cm}
\hspace*{-3.03mm} %
\includegraphics[width=6.6cm,clip]{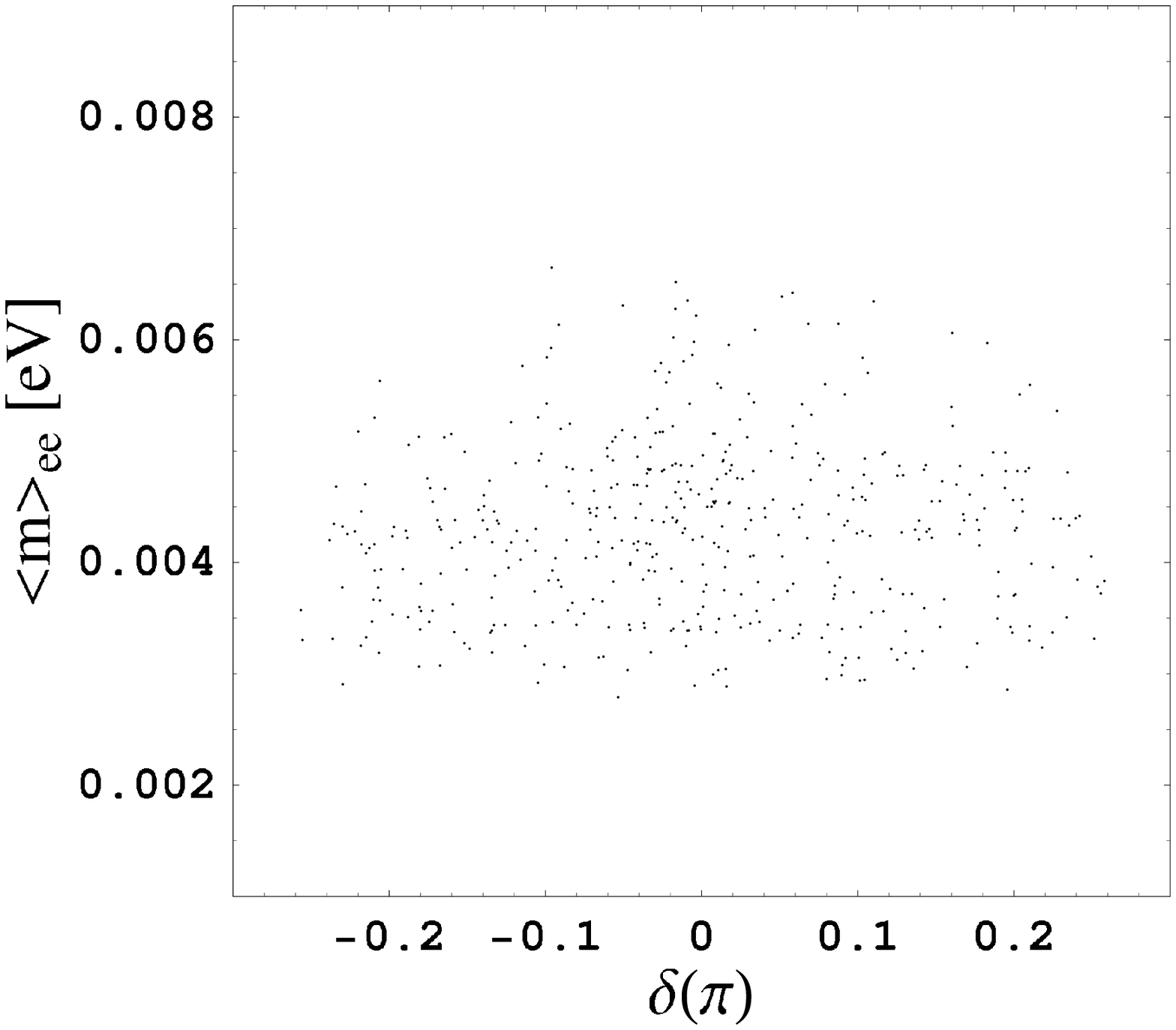}
\end{center}
%%%------------%%%
\caption{The allowed regions of $\sin^2 \theta^{}_{12}$ vs $\sin^2
\theta^{}_{23}$, $R^{}_{\nu}$ vs $\sin^2 \theta^{}_{13}$ and
$\delta$ vs $\langle m \rangle^{}_{ee}$ in the $\delta^{}_{12} =0$
(original FTY) case.}
\end{figure}

%%%%%%%%%%%%%%%%%%%
% Fig. 2
%%%%%%%%%%%%%%%%%%%
\begin{figure}
\begin{center}
\includegraphics[width=6.5cm,clip]{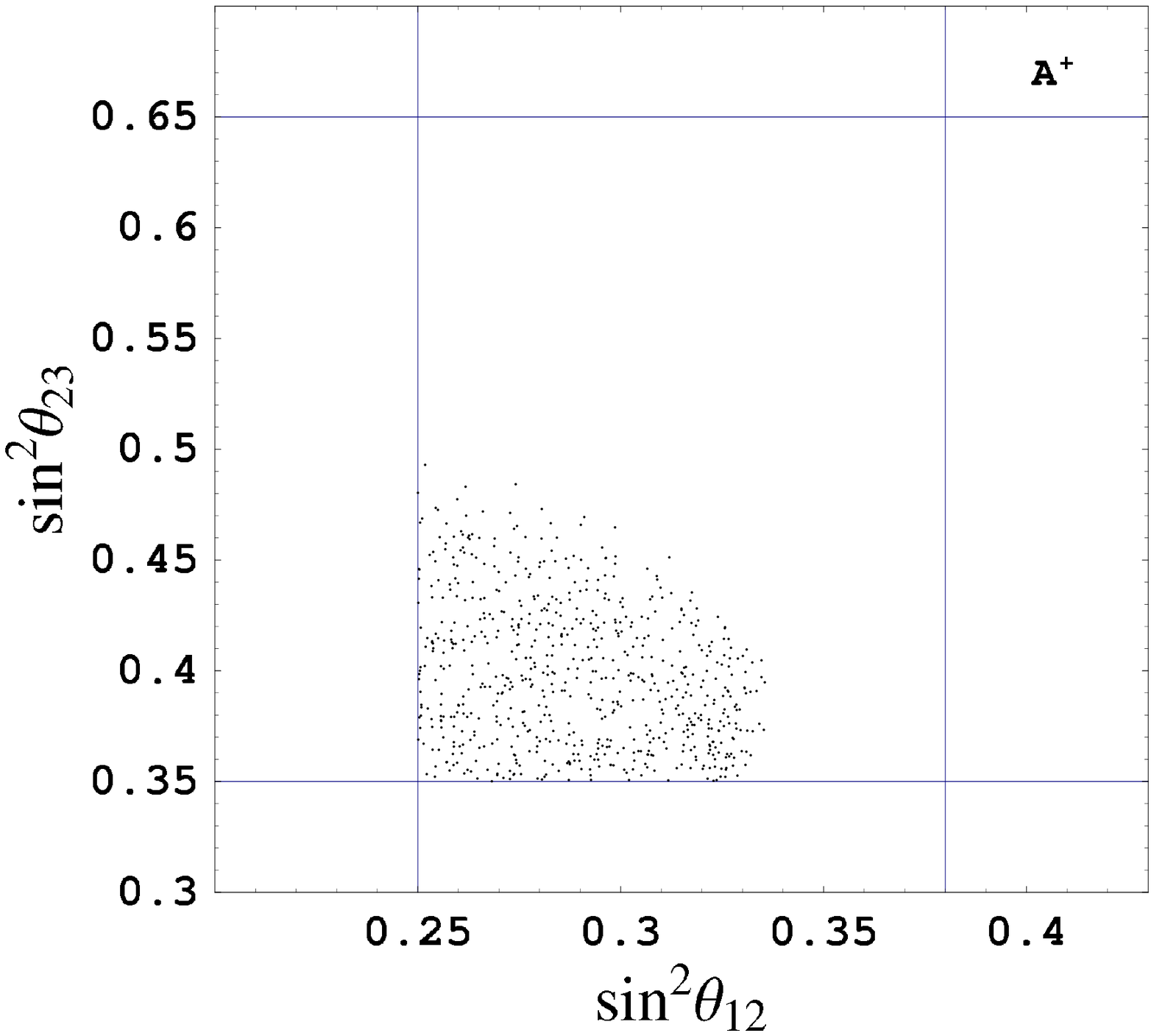}
{\hspace{0.2cm}}
\includegraphics[width=6.5cm,clip]{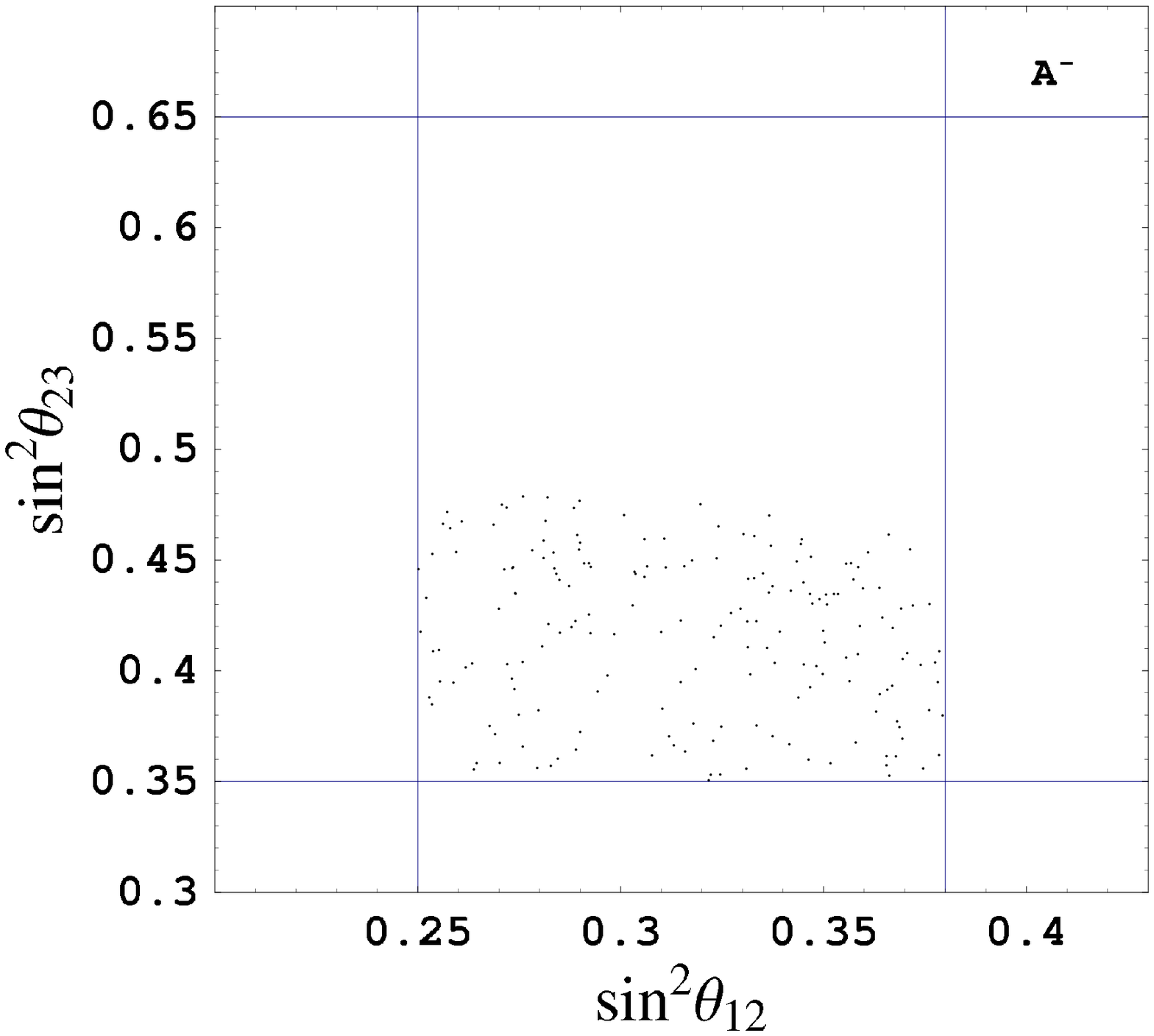}
\\
\vspace{0.4cm}
\hspace*{-1.5mm} %
\includegraphics[width=6.5cm,clip]{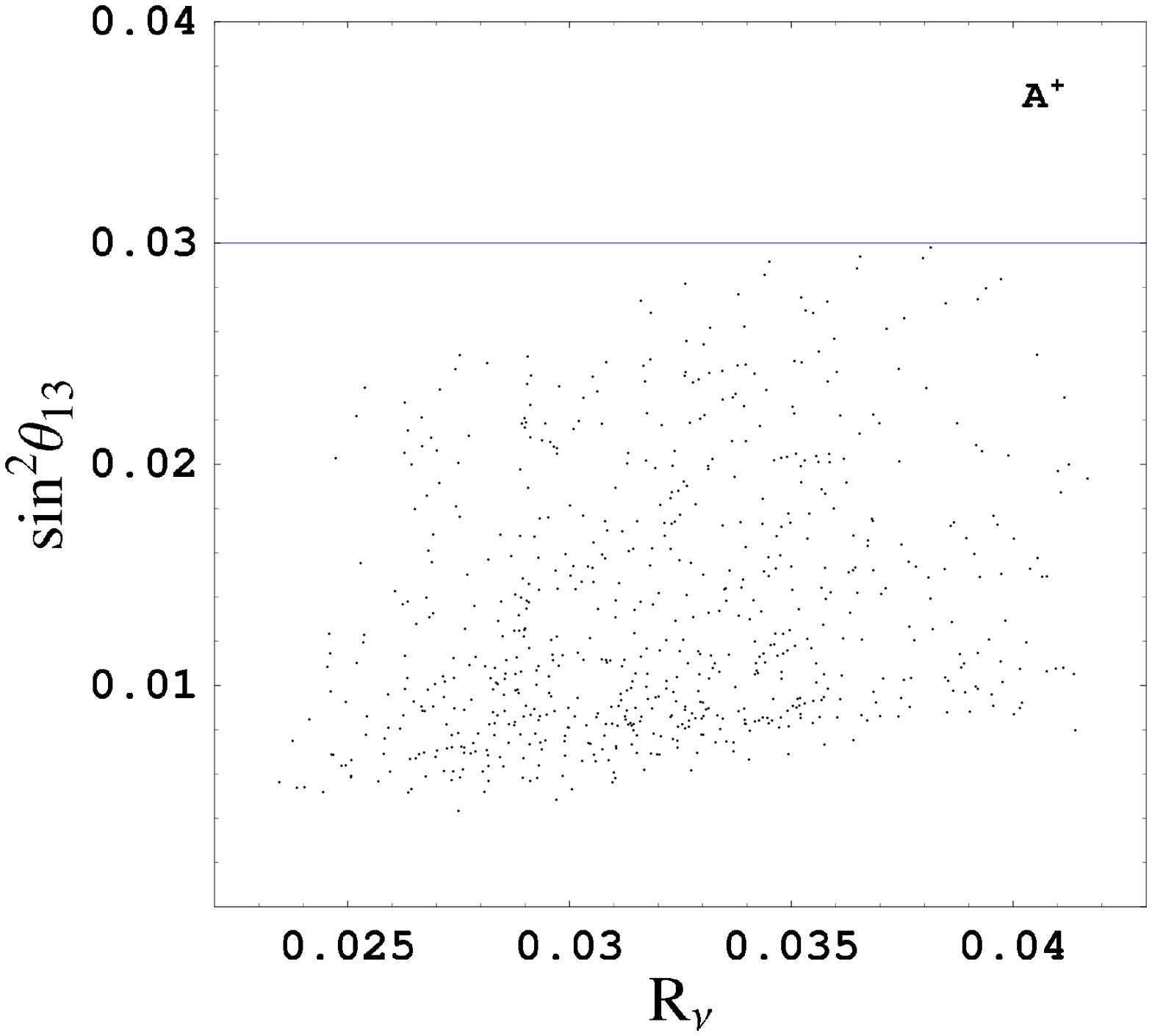}
{\hspace{0.2cm}}
\includegraphics[width=6.5cm,clip]{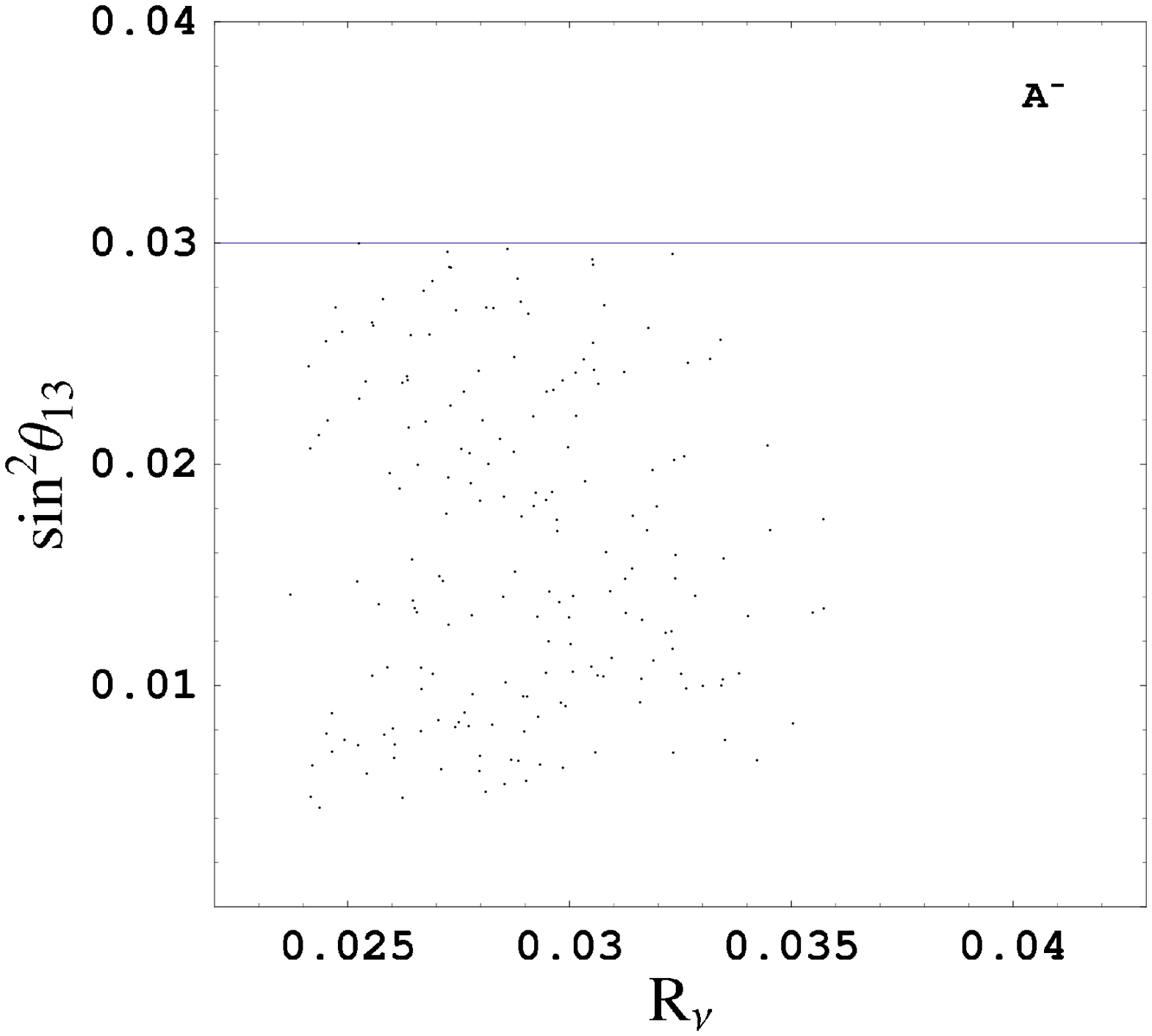}
\\
\vspace{0.4cm}
\hspace*{-3.03mm} %
\includegraphics[width=6.6cm,clip]{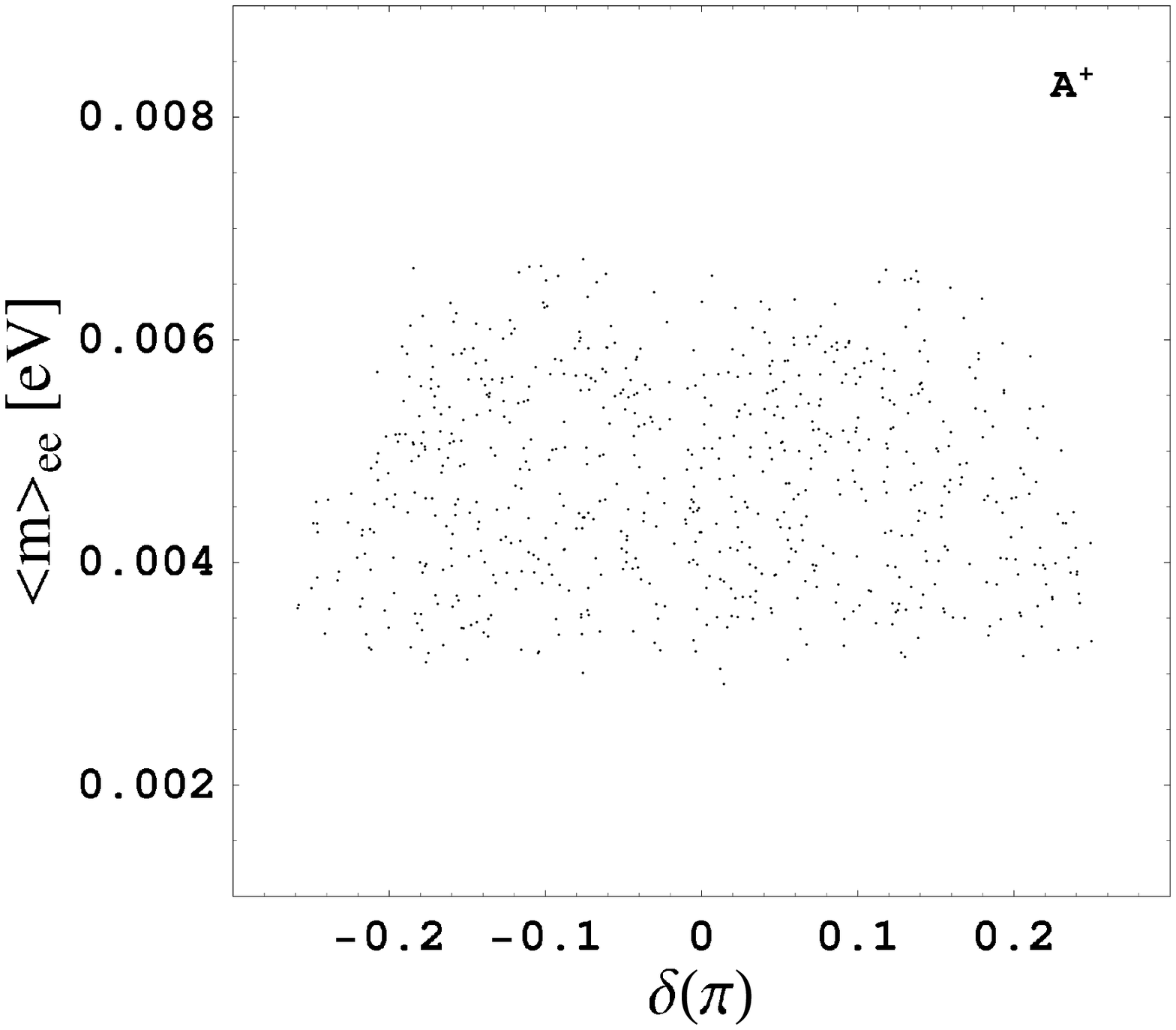}
{\hspace{0.1cm}}
\includegraphics[width=6.6cm,clip]{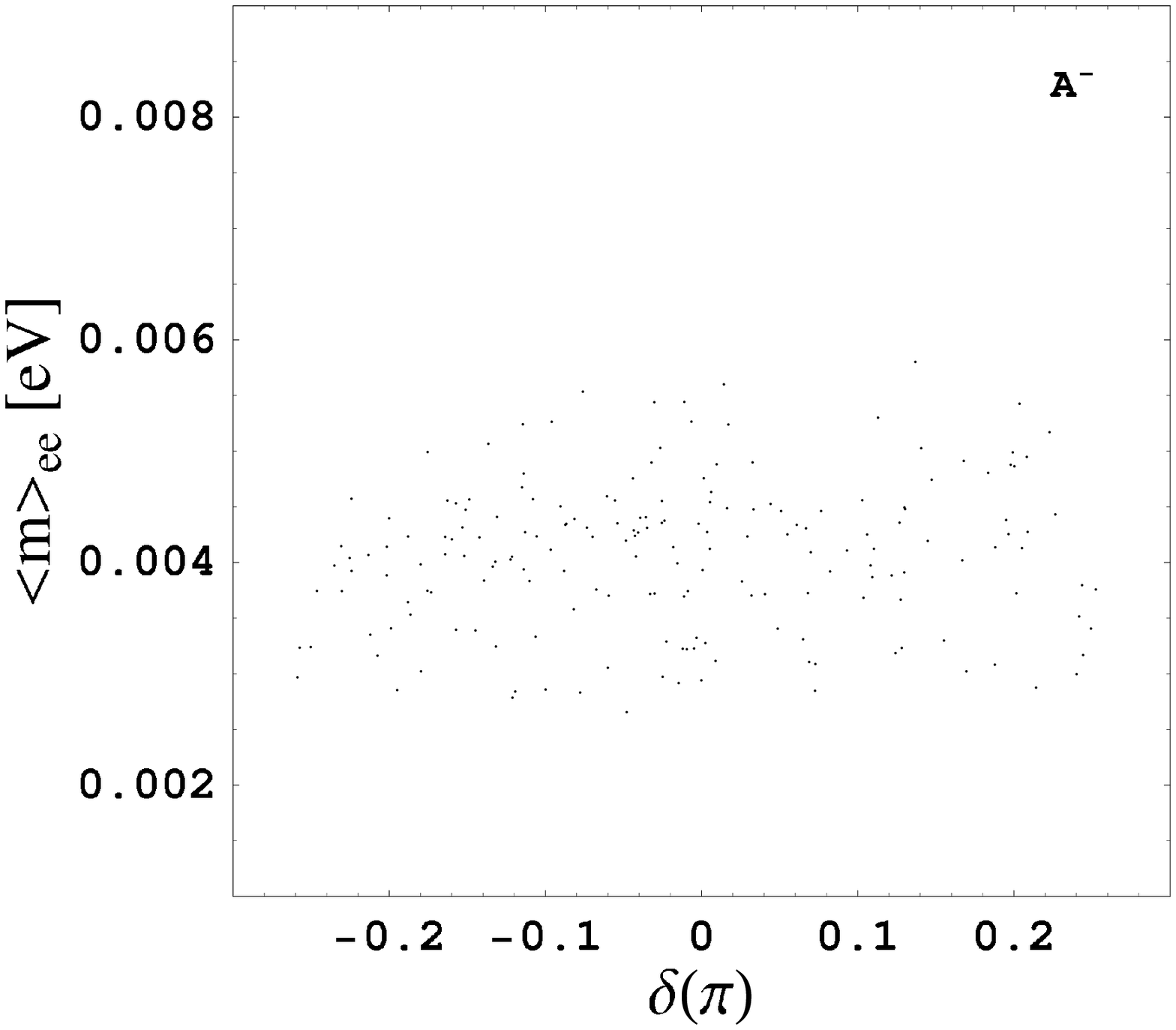}
\end{center}
%%%------------%%%
\caption{The allowed regions of $\sin^2 \theta^{}_{12}$ vs $\sin^2
\theta^{}_{23}$, $R^{}_{\nu}$ vs $\sin^2 \theta^{}_{13}$ and
$\delta$ vs $\langle m \rangle^{}_{ee}$ in cases (A$^+$) and
(A$^-$) with $|\delta^{}_{12}| =0.25$.}
\end{figure}

%%%%%%%%%%%%%%%%%%%
% Fig. 3
%%%%%%%%%%%%%%%%%%%
\begin{figure}
\begin{center}
\includegraphics[width=6.5cm,clip]{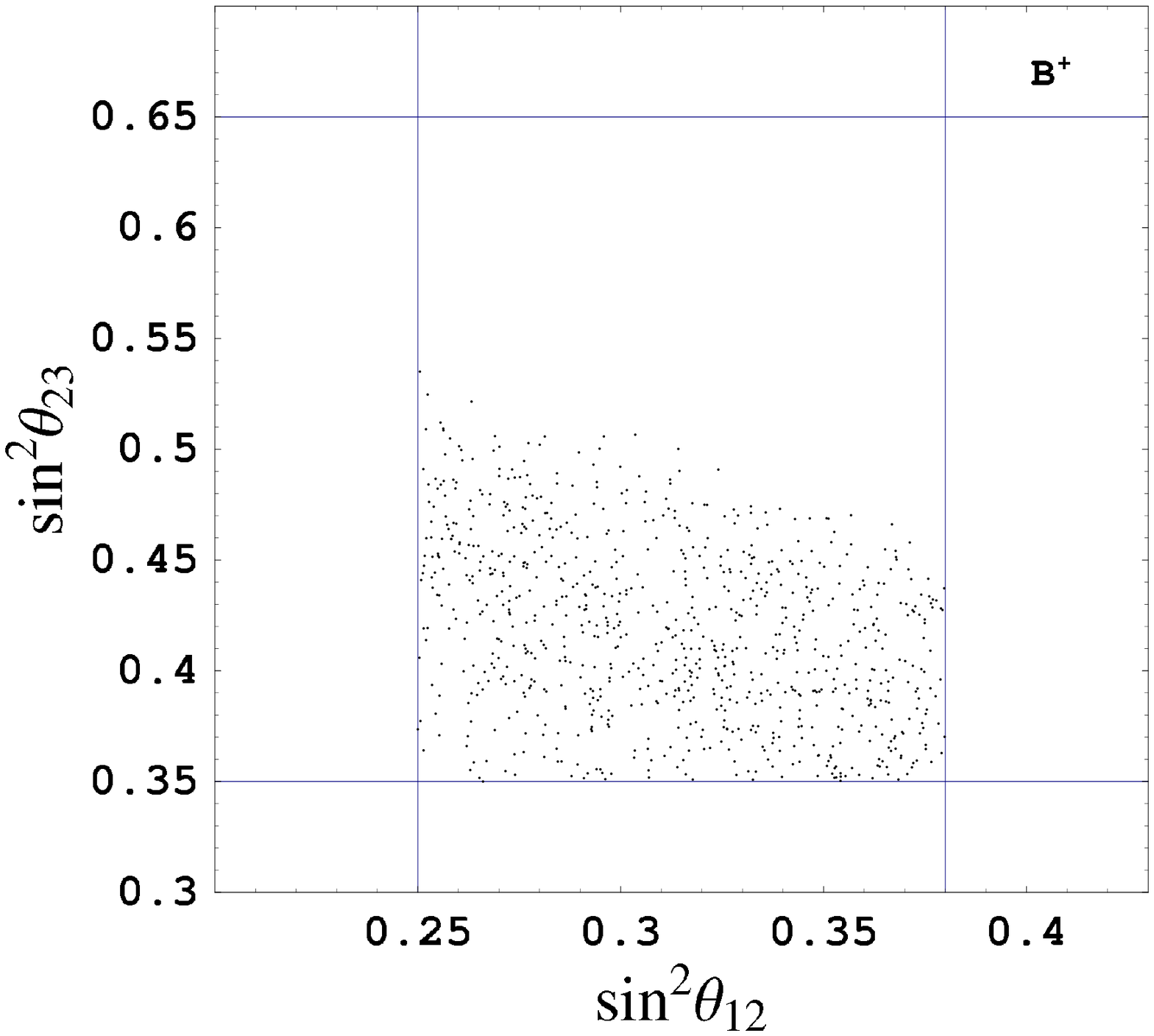}
{\hspace{0.2cm}}
\includegraphics[width=6.5cm,clip]{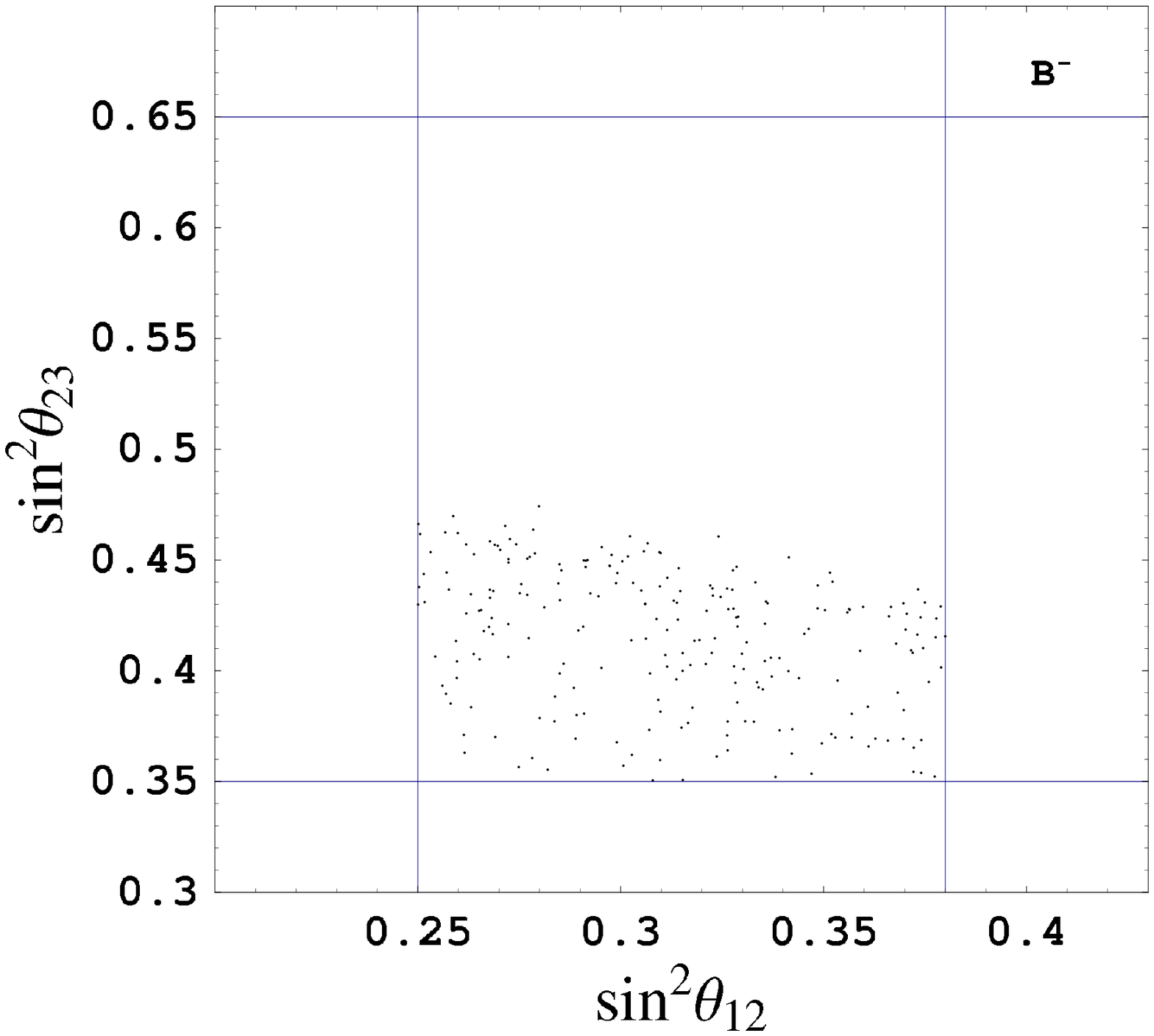}
\\
\vspace{0.4cm}
\hspace*{-1.5mm} %
\includegraphics[width=6.5cm,clip]{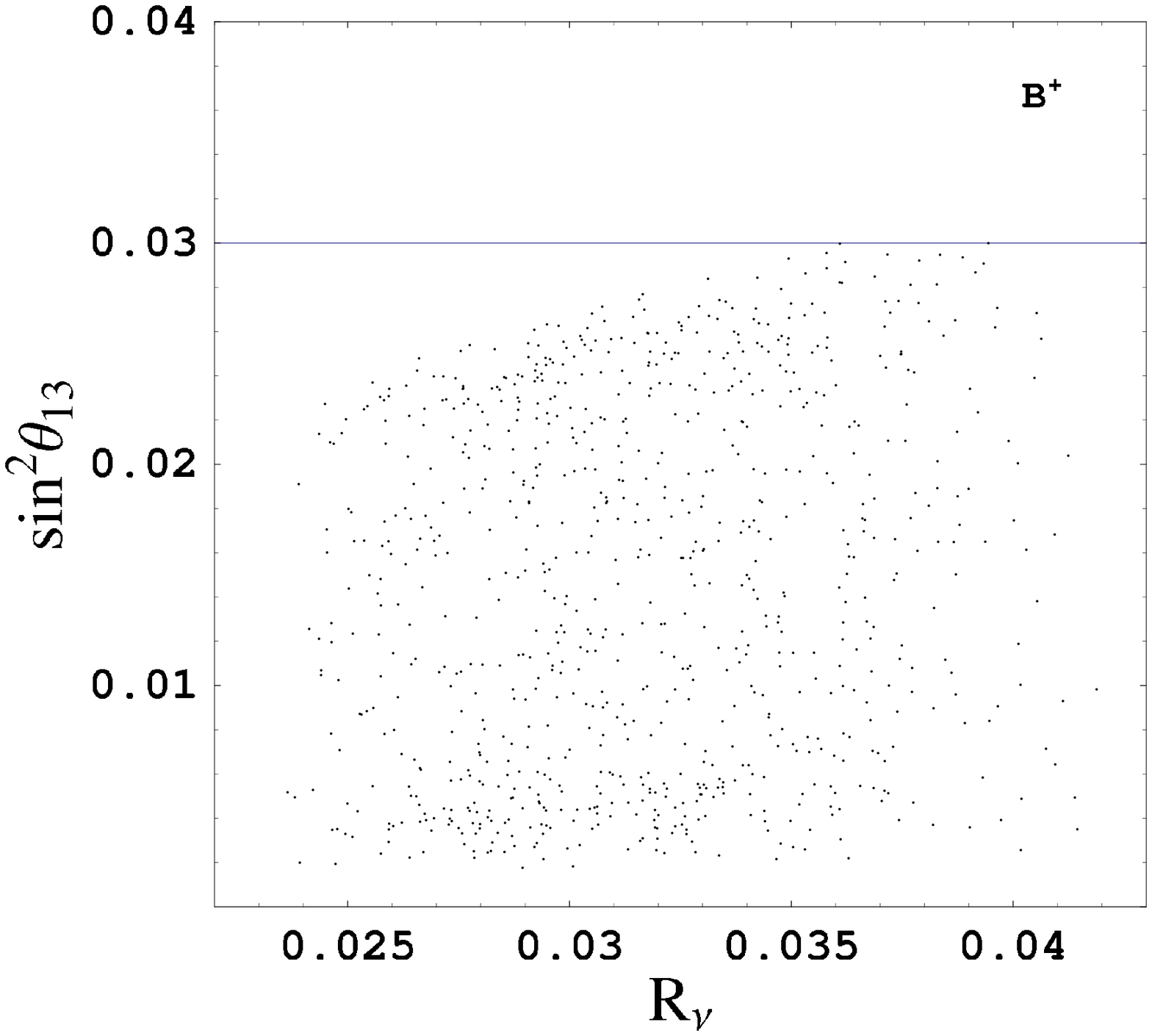}
{\hspace{0.2cm}}
\includegraphics[width=6.5cm,clip]{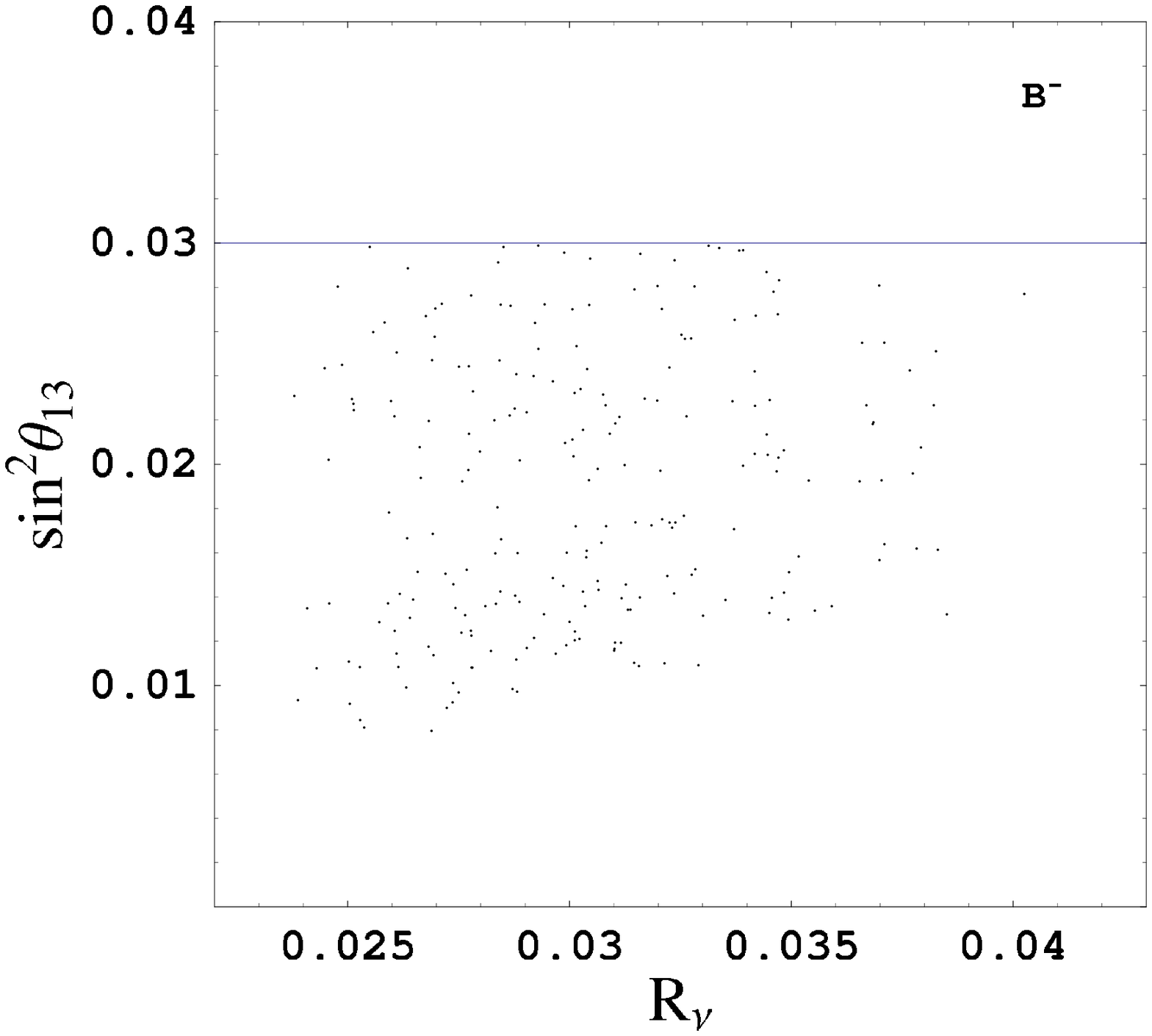}
\\
\vspace{0.4cm}
\hspace*{-3.03mm} %
\includegraphics[width=6.6cm,clip]{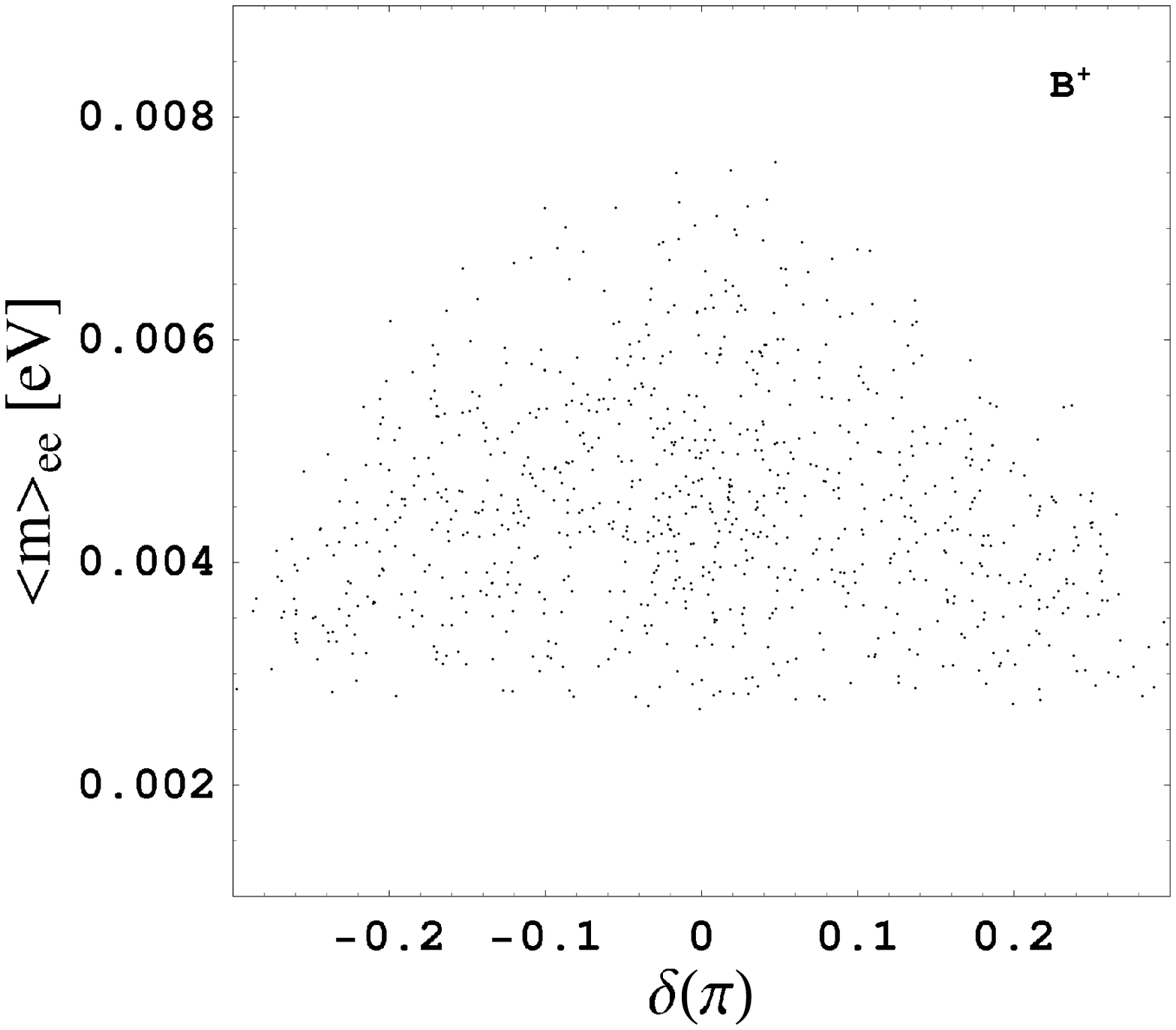}
{\hspace{0.1cm}}
\includegraphics[width=6.6cm,clip]{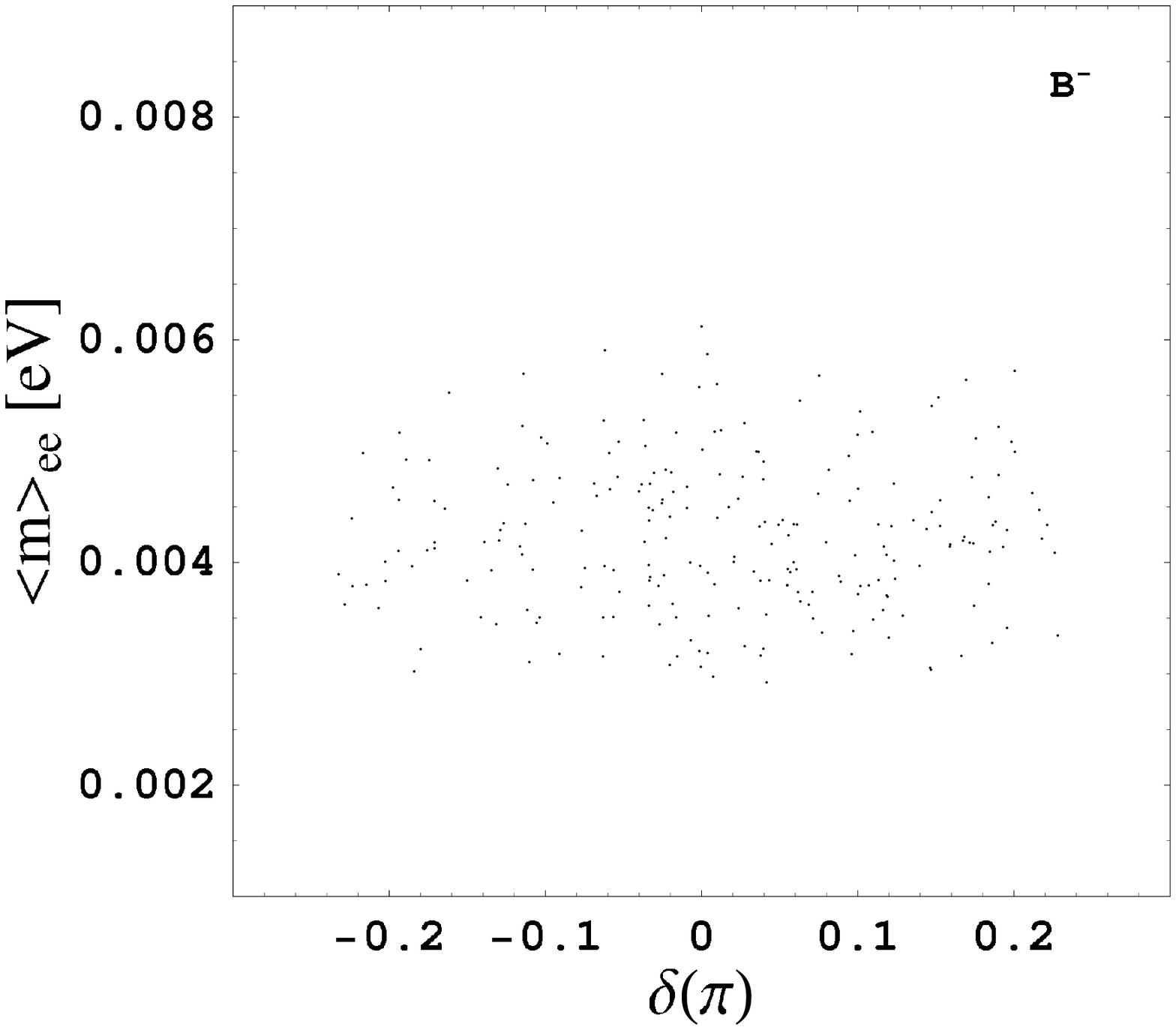}
\end{center}
%%%------------%%%
\caption{The allowed regions of $\sin^2 \theta^{}_{12}$ vs $\sin^2
\theta^{}_{23}$, $R^{}_{\nu}$ vs $\sin^2 \theta^{}_{13}$ and
$\delta$ vs $\langle m \rangle^{}_{ee}$ in cases (B$^+$) and
(B$^-$) with $|\delta^{}_{12}| =0.25$.}
\end{figure}

%%%%%%%%%%%%%%%%%%%
% Fig. 4
%%%%%%%%%%%%%%%%%%%
\begin{figure}
\begin{center}
\includegraphics[width=6.5cm,clip]{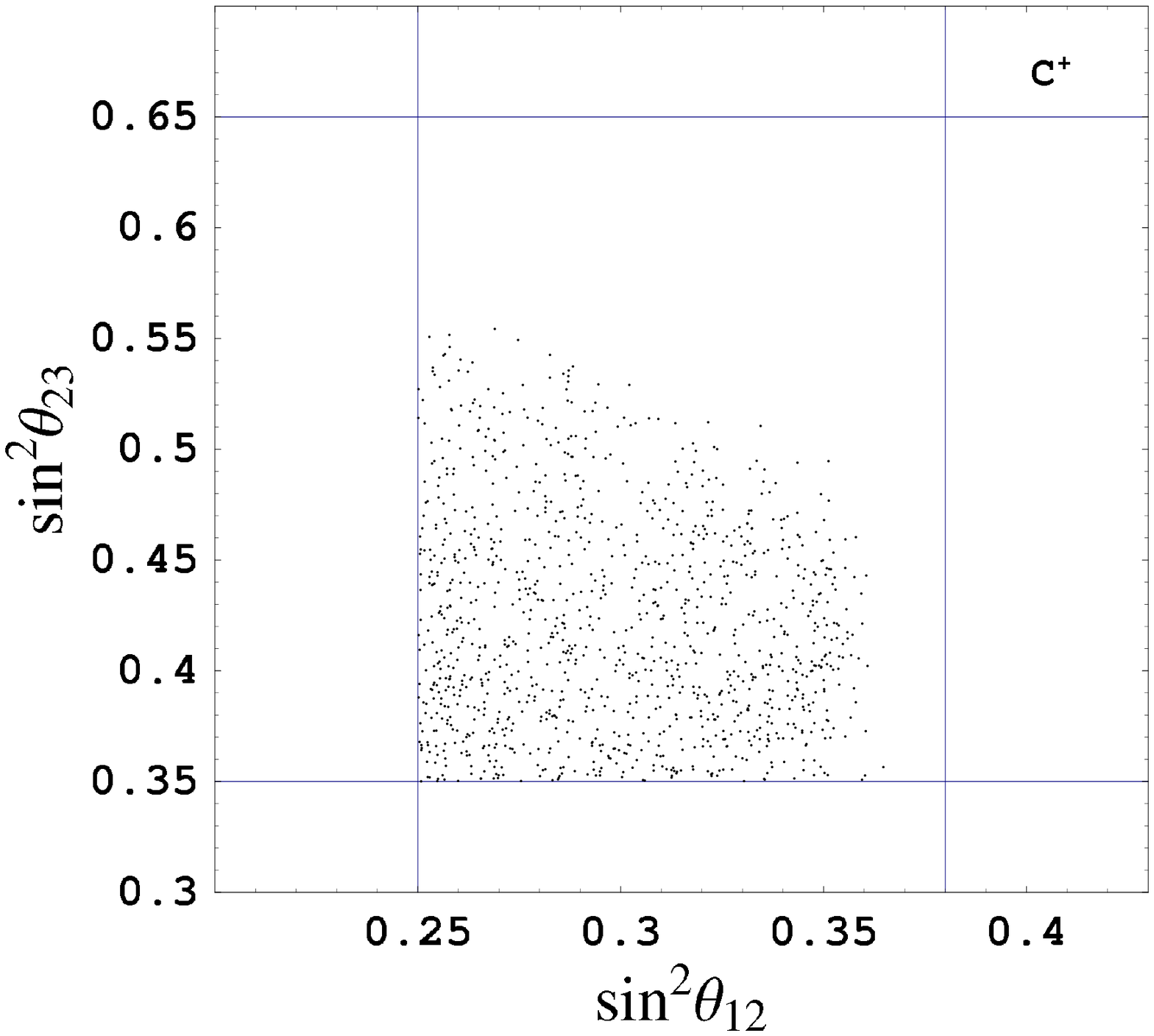}
{\hspace{0.2cm}}
\includegraphics[width=6.5cm,clip]{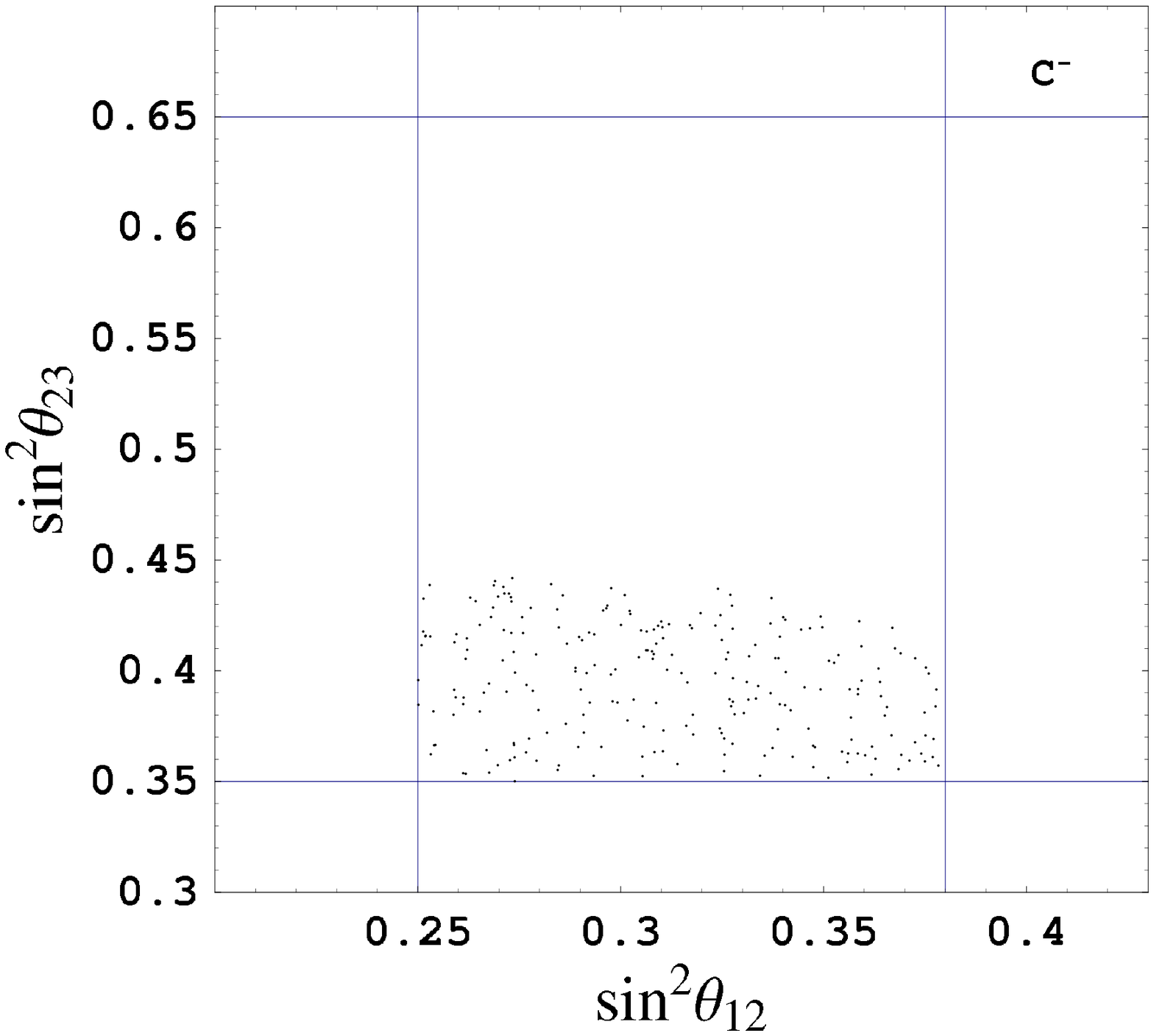}
\\
\vspace{0.4cm}
\hspace*{-1.5mm} %
\includegraphics[width=6.5cm,clip]{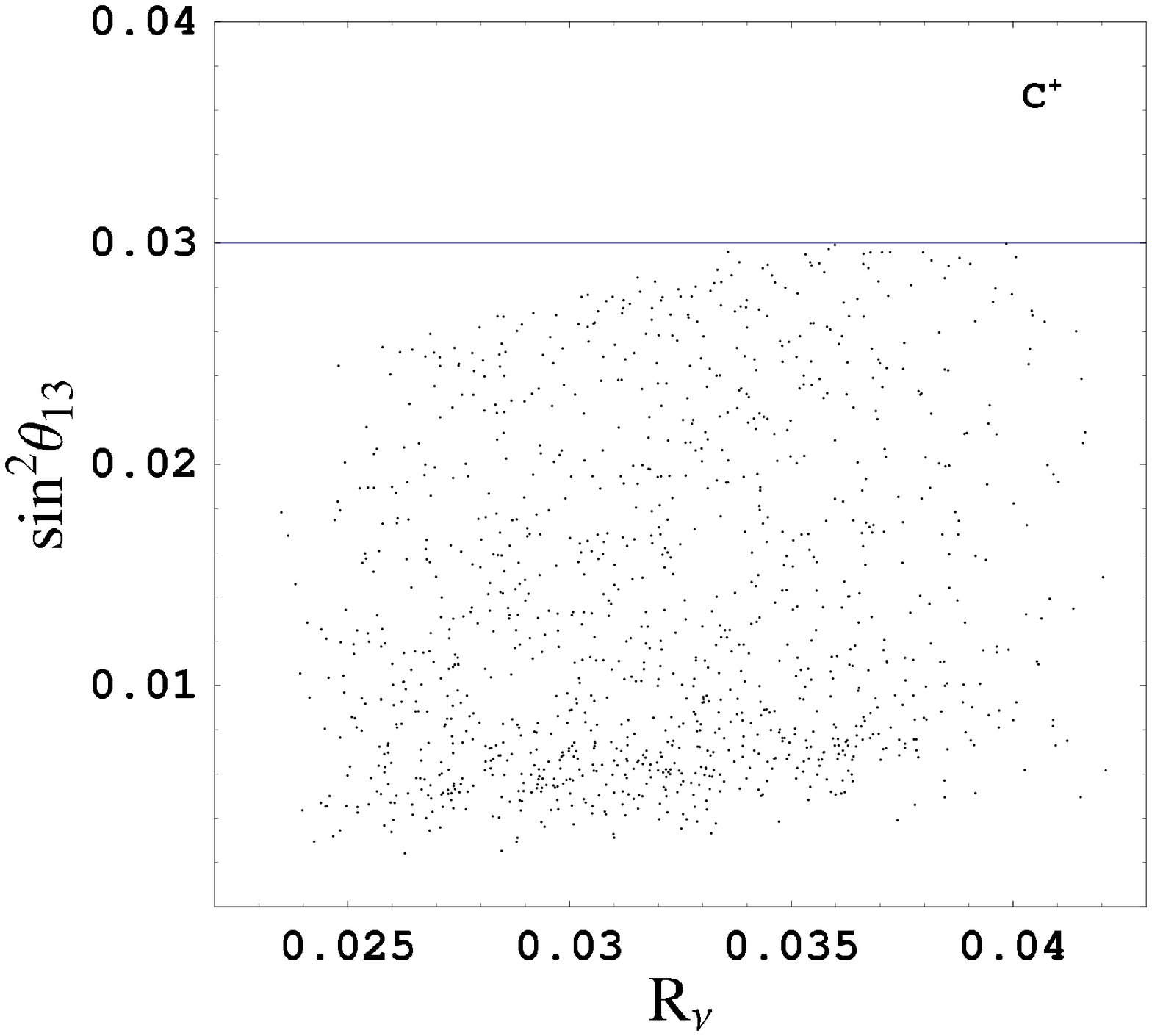}
{\hspace{0.2cm}}
\includegraphics[width=6.5cm,clip]{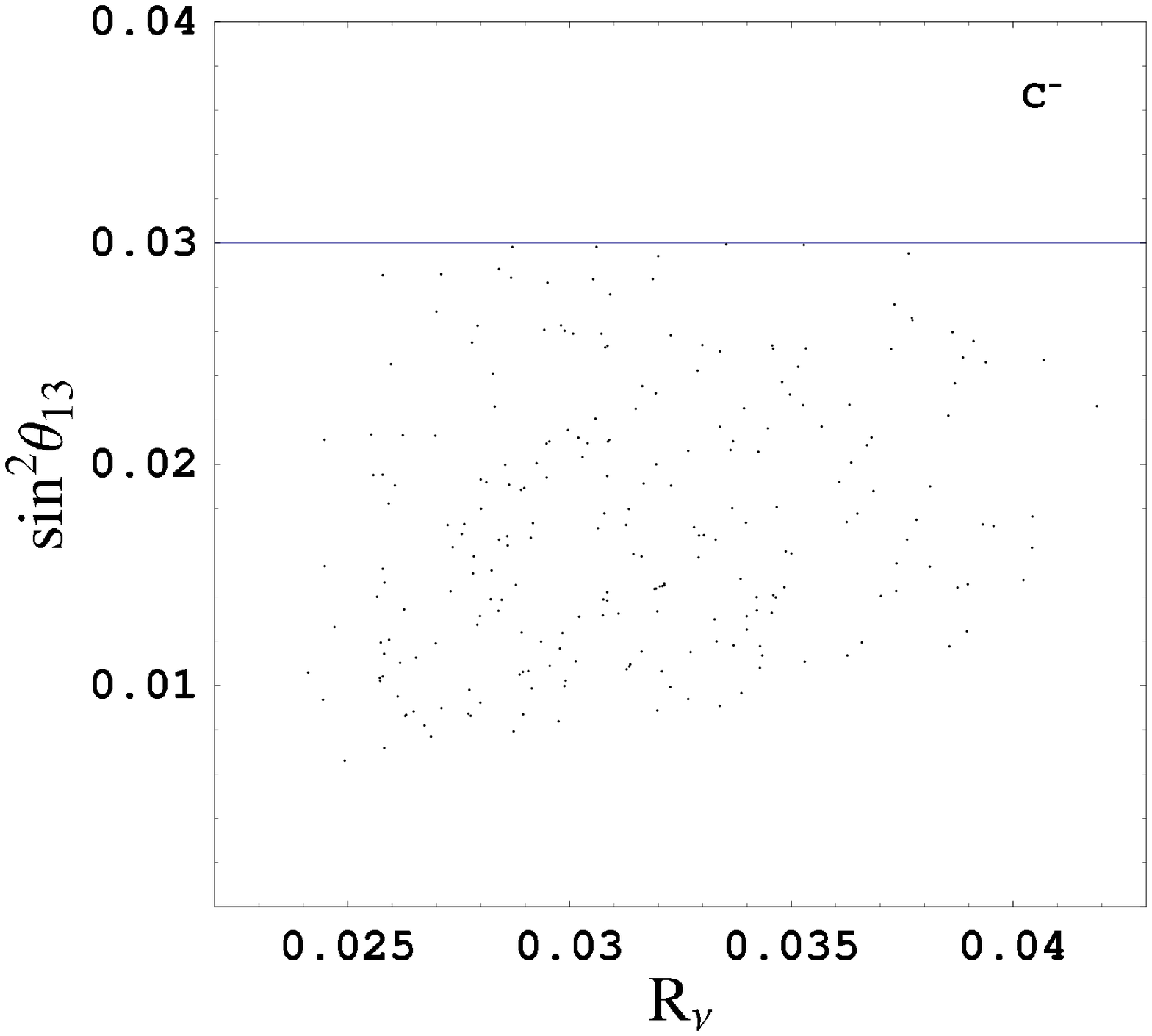}
\\
\vspace{0.4cm}
\hspace*{-3.03mm} %
\includegraphics[width=6.6cm,clip]{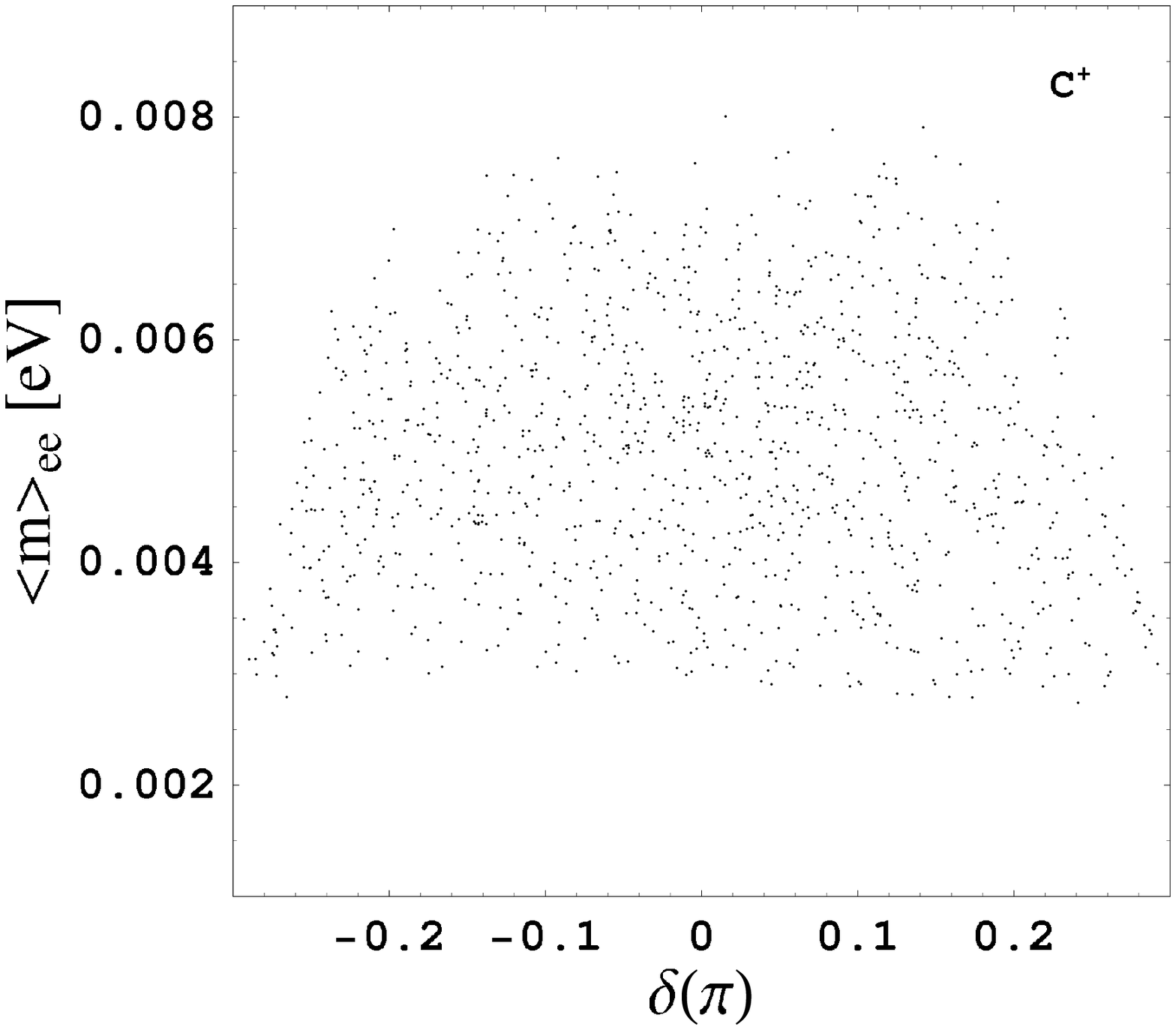}
{\hspace{0.1cm}}
\includegraphics[width=6.6cm,clip]{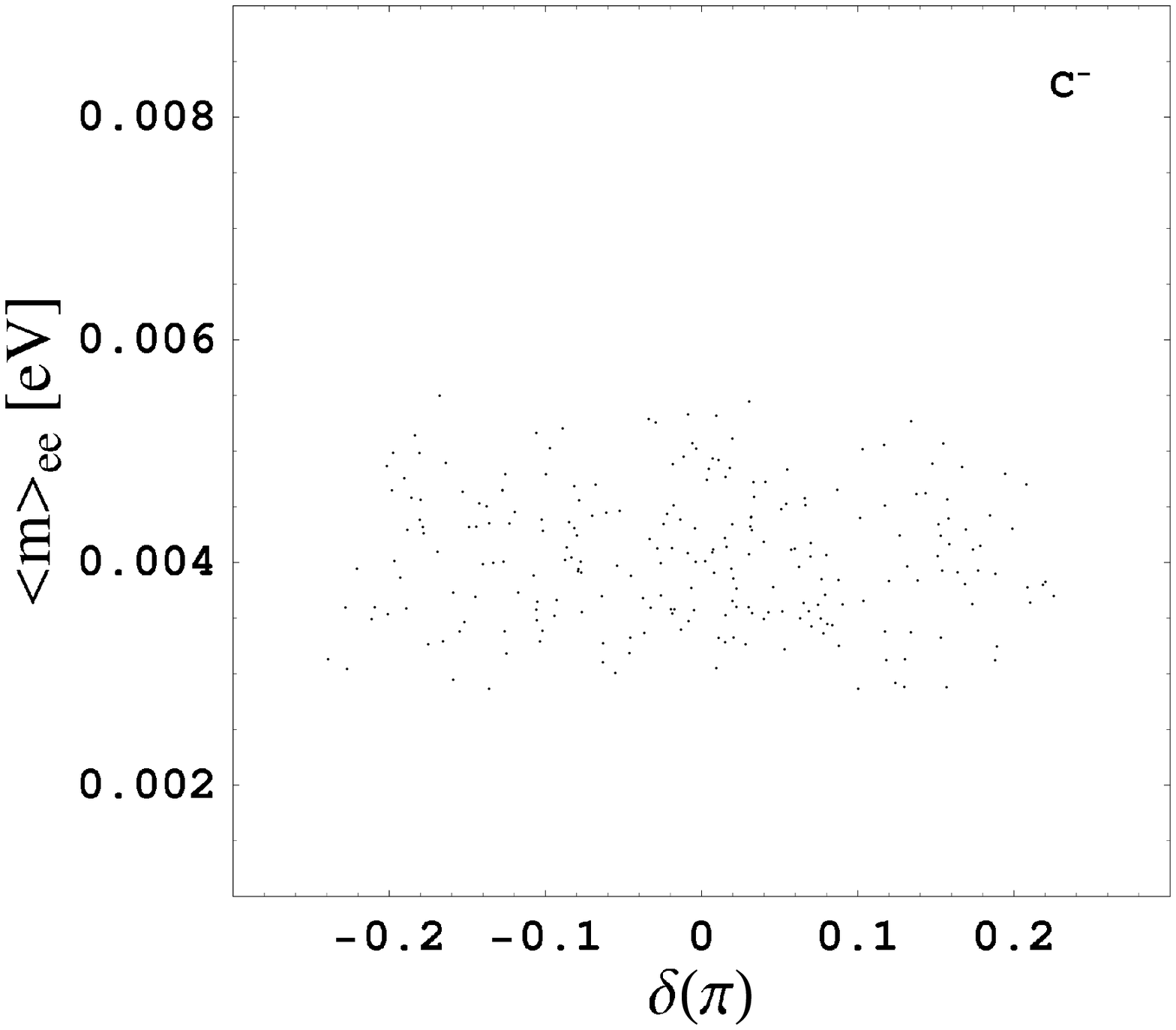}
\end{center}
%%%------------%%%
\caption{The allowed regions of $\sin^2 \theta^{}_{12}$ vs $\sin^2
\theta^{}_{23}$, $R^{}_{\nu}$ vs $\sin^2 \theta^{}_{13}$ and
$\delta$ vs $\langle m \rangle^{}_{ee}$ in cases (C$^+$) and
(C$^-$) with $|\delta^{}_{12}| =0.25$.}
\end{figure}

\end{document}